# OSIRIS-REx Lessons Learned Report

Insights from the Journey to Bennu and Back

**OSIRIS-REx Final Mission Report**

Prepared by Dante S. Lauretta, Heather L. Enos, and C.W.V. Wolner

# Abstract

OSIRIS-REx, NASA's third New Frontiers mission, was the first U.S. mission to return a sample from an asteroid. Proposed in 2010 for an $811 million lifecycle cost, the mission was completed for $828 million, returned more than $52 million of Headquarters reserves unused, and delivered at least 121.6 g of pristine material from near-Earth asteroid (101955) Bennu in September 2023, more than twice the Level-1 requirement. It did so despite a surface far more rugged than predicted, the discovery of particle ejection, sampling in a regolith much weaker than expected, operations during the COVID-19 pandemic, and an off-nominal parachute sequence on Earth return. This report sets out the lessons learned by the mission's leadership across six areas: leadership and governance; cost, schedule and risk management; systems engineering; mission operations; sample return, curation and analysis; and data archiving. Recurring themes are that a PI-led structure integrating science, engineering and project management gives the discipline to make fast, informed decisions; that reserves and schedule margin spent early to retire risk, such as funding Natural Feature Tracking as a backup navigation system, buy the flexibility to absorb discovery; that heritage hardware reduces risk only when its verification and engineering knowledge are preserved; and that contamination knowledge, sample science and data stewardship are mission-long responsibilities, not post-flight tasks. Each section pairs background with specific recommendations for future sample-return and small-body missions.

# Prologue

## Why should someone who wasn't on OSIRIS-REx read this report?

In 2010, we proposed a mission to return **at least 60 grams** from a primitive asteroid that had never been visited. At the time, Bennu was still known as **1999 RQ36**, and we believed it would be a relatively benign, smooth object (Lauretta et al., 2019a). Almost every major assumption about the sampling event proved incomplete or wrong.

- Bennu’s surface is rugged and dominated by boulders (DellaGiustina et al., 2019; Lauretta et al., 2019a; Walsh et al., 2019).
- The navigation problem was far harder than anticipated because sampleable material was confined to areas that were very small relative to the expected accuracy of the guidance system at launch (Lauretta et al., 2019a, 2021).
- A back-up autonomous navigation system had to become operational to safely contact the surface with far less margin than anticipated (Lauretta et al., 2021; Norman et al., 2022).

- The spacecraft encountered particle ejection events never before observed (Lauretta et al., 2019b).
- COVID-19 struck during the most critical operational period (Burns & Lauretta, 2025; Getzandanner et al., 2022).
- The sampler head jammed after collecting more material than expected as it sunk into the surprisingly weak regolith (Burns & Lauretta, 2025; Lauretta et al., 2022; Lauretta, Connolly, et al., 2024).
- The drogue parachute deployed under off-nominal conditions during Earth return (Lauretta, Connolly, et al., 2024; NASA, 2023; Williams et al., 2025).

Yet despite all of this, the mission not only succeeded, it substantially exceeded every Level-1 requirement.

These are the lessons from that journey.

## Building a Mission Designed to Learn

In May 2011, NASA selected OSIRIS-REx as the third New Frontiers mission with an ambitious objective: return a pristine sample from the primitive carbonaceous asteroid (101955) Bennu (Lauretta et al., 2017). The mission sought to answer fundamental questions about the formation of the solar system, the origin of organic matter, the delivery of life's essential ingredients to Earth, and the evolution of potentially hazardous asteroids. Success required returning at least 60 grams of uncontaminated regolith while simultaneously conducting a comprehensive geological, geochemical, and dynamical investigation of an asteroid never before explored (Lauretta et al., 2017).

From the beginning, the project recognized that achieving these scientific objectives would require more than innovative engineering. Sample-return missions combine nearly every discipline of planetary exploration: spacecraft development, precision navigation, autonomous operations, contamination control, sample curation, and decades of scientific investigation. Success would depend not only on technical excellence, but on disciplined management, integrated systems engineering, realistic cost and schedule planning, and a culture capable of adapting to discoveries that could not be anticipated during formulation.

During the mission selection process, the project identified the challenges that would determine success. The team focused on four overarching risks: executing the unprecedented operations required for asteroid sample return; preventing requirements creep; maintaining leadership continuity throughout a mission expected to span more than two decades; and establishing a cost and schedule baseline that was credible and executable. These concerns shaped every aspect of the project's management philosophy.

Detailed design reference missions, rigorous requirements management, integrated planning and control, comprehensive cost estimating, disciplined reserve management, and succession planning were incorporated from the outset—not as administrative exercises, but as mission-enabling capabilities.

The outcome demonstrated the value of that approach. In 2010, OSIRIS-REx was proposed with a total lifecycle cost of **$811 million**, while NASA Headquarters retained **$70 million** in Unallocated Future Expenses as additional programmatic reserve. In spite of numerous unexpected challenges during the Bennu encounter, the mission completed development and operations for **$828 million**, returning more than **$52 million** of Headquarters reserves unused. It returned **121.6 grams** of pristine Bennu material—more than twice the Level-1 requirement (Burns & Lauretta, 2025; Lauretta, Connolly, et al., 2024)—and has already enabled more than **200 peer-reviewed scientific publications**. Every Level-1 mission requirement was successfully met or exceeded.

This report documents the management, engineering, operational, and scientific lessons in the process. It is not intended as a chronological history of OSIRIS-REx, nor as a catalog of isolated successes and failures. Rather, it captures the principles that allowed the project to remain aligned with its objectives while adapting to uncertainty over almost 20 years of formulation, development, operations, sample return, curation, and scientific investigation.

The central lesson is simple: **Successful exploration missions are not those that avoid surprises—they are those that are deliberately designed to adapt to them.**

# 1. Leadership and Governance

## 1.1 PI-Led Mission Management

**Key Theme:** A PI-led management structure with integrated planning and control promotes alignment between science objectives, technical execution, cost, schedule, and risk.

### Background

The PI-led management structure was a fundamental contributor to the success of OSIRIS-REx. From the outset, the mission integrated scientific leadership with project planning, systems engineering, cost, schedule, risk management, and technical execution, ensuring that all major decisions were evaluated against the mission's primary objective of safely acquiring and returning a pristine sample from asteroid Bennu.

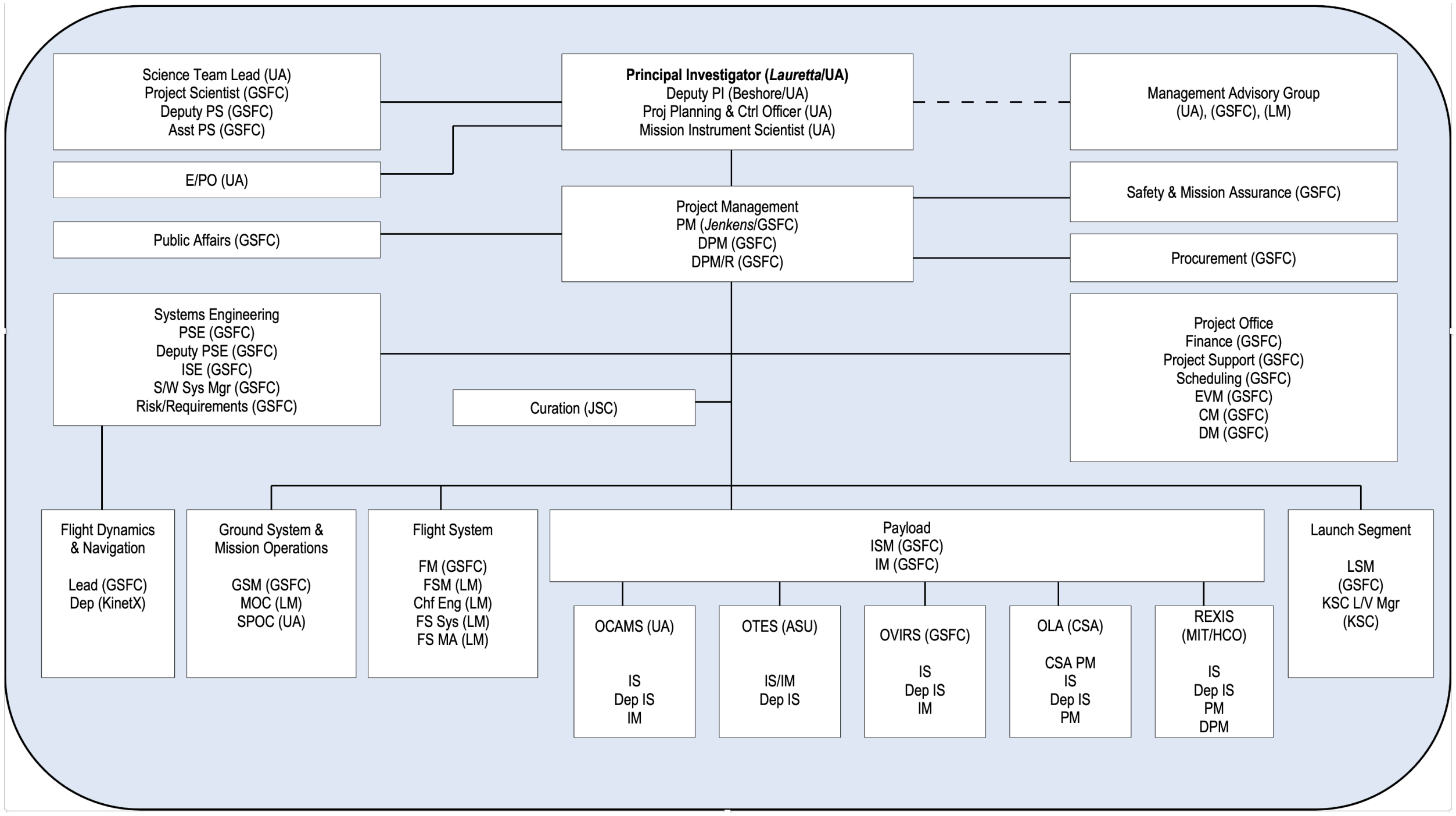

**Figure 1**. OSIRIS-REx Project Organization Chart during development. Credit: OSIRIS-REx Program-level Requirements Appendix (PLRA).

A defining strength of the PI-led model was the establishment of a unified management structure. Rather than operating as parallel organizations, the science, engineering, project management, and systems engineering teams functioned within a single decision-making framework. This integration enabled rapid trade studies, efficient decision-making, and consistent evaluation of technical, scientific, cost, and schedule impacts against a common set of mission objectives.

The mission also benefited from an experienced partnership among the University of Arizona, NASA Goddard Space Flight Center, and Lockheed Martin. These organizations had worked together in various configurations on previous missions, allowing the project to begin with a high level of mutual trust, established working relationships, and organizational maturity.

Mission management remained rigorously requirements-driven throughout development. Beginning in Phase A, detailed Statements of Work (SOWs) and Basis of Estimates (BOEs) were developed to align technical scope, staffing, schedule, and budget with the Design Reference Mission and Level-1 requirements. These planning products were treated as living documents and reviewed regularly as the mission evolved. Throughout the project, proposed additions were evaluated against a simple guiding question: "Is this a requirement or a desirement?" This disciplined approach prevented requirements creep and maintained focus on mission success.

A dedicated Project Planning and Control Office within the PI Office maintained integrated visibility into budget, schedule, staffing, technical performance, and mission risk. Working closely with science, engineering, and project management, this organization ensured that planning remained realistic, risks were tracked continuously, and resources remained aligned with mission priorities. By integrating planning, execution, and performance assessment within a single organization, the project maintained transparency, accountability, and the ability to adapt without compromising mission objectives.

The PI also maintained a sustained presence at both NASA Goddard Space Flight Center and NASA Headquarters during development. Regular face-to-face engagement strengthened communication across institutional boundaries, built mutual trust among mission partners, and accelerated resolution of technical and programmatic issues. For a complex, multi-institution mission, this level of personal engagement proved to be a critical investment in mission cohesion.

## Lessons Learned

- **Establish a PI-led management structure** that integrates scientific leadership with project planning, systems engineering, cost, schedule, and risk management under a unified decision-making framework.
- **Embed a dedicated Project Planning and Control Office** within the PI organization to maintain continuous visibility into budget, schedule, technical performance, staffing, and mission risk, enabling timely, transparent, and informed decision-making throughout the mission lifecycle.
- **Maintain disciplined requirements management** throughout the mission lifecycle. Additional capability should be incorporated only when it directly advances the mission's primary objectives; more is not always better. Throughout OSIRIS-REx, maintaining focus on successful sample return prevented unnecessary scope growth from increasing mission risk.
- **Evaluate all proposed scope changes using objective criteria**—including the guiding question, "Is this a requirement or a desirement?"—to prevent requirements creep while maintaining alignment with mission objectives.
- **Invest in regular face-to-face engagement** among the Principal Investigator, project leadership, NASA centers, contractors, and Headquarters to strengthen trust, improve transparency, and accelerate resolution of technical and programmatic issues.

---

## 1.2 Management Advisory Council

**Key Theme:** Senior institutional leadership provides continuity, strategic oversight, and conflict resolution beyond the authority of the project.

### Background

OSIRIS-REx established a Management Advisory Council (MAC) composed of senior executives from each of the principal partner institutions, including the President of the University of Arizona, the Director of NASA Goddard Space Flight Center, and the Vice President of Lockheed Martin Space Systems. The MAC served as the highest-level governance body external to the project, providing strategic oversight and institutional support when issues extended beyond the authority of the project management team.

The Council's charter included facilitating access to personnel, facilities, and procurement resources; resolving inter-organizational conflicts; aligning institutional priorities; and responding to major programmatic risks or unforeseen events. Because its members represented the senior leadership of each partner organization, the MAC was uniquely positioned to resolve issues that could not be addressed through normal project management channels.

The value of the MAC became particularly evident during significant leadership transitions, including the transfer of Principal Investigator (PI) responsibilities following the passing of the originating PI and the subsequent appointments of a new PI and Deputy PI. These transitions could have disrupted project execution or weakened institutional alignment. Instead, the MAC provided continuity of governance, reinforced organizational commitment, and maintained strategic alignment among the University of Arizona, NASA, and Lockheed Martin throughout the transition.

Beyond leadership transitions, the existence of an engaged executive governance body provided the project with an established mechanism for resolving institutional issues before they affected technical execution, enabling the project leadership to remain focused on mission success.

### Lessons Learned

- **Establish a Management Advisory Council early** in the mission, composed of senior leadership from each principal partner organization, to resolve institutional issues, align organizational priorities, facilitate access to critical resources, maintain executive-level communication across participating organizations, and provide strategic oversight beyond the authority of the project management team.

- **Maintain active engagement** throughout the mission lifecycle rather than convening the Council only during crises. An established governance structure is most effective when relationships and decision pathways are developed before major issues arise.
- **Leverage the MAC to preserve continuity** during leadership transitions, organizational changes, and unexpected mission events, ensuring that institutional commitment and strategic alignment remain intact while project leadership focuses on mission execution.

---

## 1.3 Leadership Continuity and Team Stability

**Key Theme:** Continuity of personnel, institutional knowledge, management processes, and mission planning preserves effectiveness throughout long-duration projects.

### Background

OSIRIS-REx benefited from exceptional continuity among its leadership and technical teams throughout formulation, development, operations, sample return, and curation. Many key personnel remained in their roles for more than a decade, preserving institutional knowledge, maintaining long-term accountability, and fostering strong working relationships across the principal partners. This continuity ensured that decisions made during formulation remained connected to their original technical rationale throughout development, operations, and sample return.

A defining strength of the mission was its end-to-end staffing philosophy. Many members of the proposal and formulation teams transitioned directly into implementation and flight operations, creating a strong sense of ownership and accountability for mission success. Likewise, involving mission operations personnel during Assembly, Test, and Launch Operations (ATLO) ensured a smooth transition into flight operations by providing early system-level knowledge, operational experience, and established working relationships. This continuity reduced learning curves, accelerated problem solving, and improved decision-making throughout the mission lifecycle.

The mission also emphasized succession planning by intentionally mentoring early-career scientists, engineers, and managers. This investment enabled leadership transitions— including the transfer of Principal Investigator responsibilities—to occur without disrupting mission objectives while preparing the next generation of planetary mission leaders.

OSIRIS-REx demonstrated that continuity extends beyond the project organization itself. Changes within supporting NASA program offices altered established communication pathways, reduced organizational efficiency, and weakened institutional memory. Long-duration missions therefore benefit from continuity not only within the project team but also across the broader programmatic organizations responsible for oversight, funding, and mission support.

Mission continuity was further strengthened through the long-term use of the Design Reference Mission (DRM) as a living systems engineering and management document. Rather than serving solely as a formulation product, the DRM evolved throughout the mission as knowledge of Bennu increased and operational assumptions changed. Its disciplined evolution maintained alignment among scientific objectives, technical requirements, schedule, cost, and risk while preserving sufficient schedule margin to accommodate new discoveries, operational refinements, and unexpected events—including Bennu's rugged terrain and the COVID-19 pandemic.

## Lessons Learned

- **Promote continuity** of key leadership, engineering, science, and project management personnel across principal partners throughout the mission lifecycle whenever practical. Long-term continuity preserves institutional knowledge, strengthens communication and accountability, reduces organizational friction, and improves decision-making efficiency.

- **Adopt a mission lifetime staffing philosophy** by involving personnel responsible for later mission phases—including operations and sample curation—early in development and system testing. Early participation improves knowledge transfer, operational readiness, and organizational ownership.

- **Plan explicitly for succession**. Leadership transitions should include deliberate knowledge transfer, clearly defined responsibilities, and governance mechanisms that preserve established management processes and institutional memory.

- **Maintain the Design Reference Mission as a living systems engineering document** throughout the mission lifecycle. As scientific understanding and operational realities evolve, mission plans should be updated while preserving alignment among mission objectives, technical requirements, cost, schedule, and risk.

- **Protect schedule margin as a strategic resource**. Realistic, risk-informed schedules provide the flexibility needed to accommodate new discoveries, evolving requirements, and unforeseen technical or operational challenges without compromising mission objectives.

---

## 1.4 Organizational Authority, Contract Alignment, and International Partnerships

**Key Theme:** Organizational authority, contractual relationships, and acquisition strategies must reinforce the intended management model and enable effective execution of mission-critical responsibilities.

### Background

A defining characteristic of OSIRIS-REx was its PI-led management structure, in which the Principal Investigator was ultimately responsible for mission success. For this model to function effectively, organizational authority, contractual relationships, and communication pathways needed to support direct coordination among the PI Office, NASA, contractors, and international partners.

During development, the project demonstrated that contractual structures can either strengthen or complicate mission management. Although the PI was responsible for overall mission success, the primary spacecraft contract was held between NASA Goddard Space Flight Center and Lockheed Martin. This traditional contractual arrangement occasionally constrained communication and decision-making by requiring interactions to follow contractual reporting relationships rather than operational needs. Although these challenges were successfully managed, the experience demonstrated that contractual authority should reinforce—not complicate—the intended management structure. When a Principal Investigator is accountable for mission success, the PI should also have clear authority and direct access to implementing organizations.

OSIRIS-REx also demonstrated the importance of matching acquisition strategy to technical and programmatic risk. The navigation laser altimeter was originally planned to be provided by an experienced aerospace contractor, but procurement ultimately shifted to a smaller organization operating under a Fixed Firm Price contract. As technical and schedule challenges emerged, the project responded by embedding personnel with the contractor, increasing management oversight, and developing Natural Feature Tracking (NFT) as an independent backup capability. These actions mitigated technical risk while preserving a critical navigation function. The experience demonstrated that higher-risk acquisition strategies require correspondingly greater investments in technical oversight, systems engineering, and contingency planning.

International partnerships presented a complementary management challenge. The OSIRIS-REx Laser Altimeter (OLA), the science instrument contributed by the Canadian Space Agency, faced funding and schedule uncertainties during formulation that made it a candidate for descoping. Rather than treating OLA solely as a contractual issue, the PI Office Team worked collaboratively with NASA, the Canadian Space Agency, MDA Space Ltd., and other partners to develop flexible implementation strategies that preserved the instrument within the project baseline. Following arrival at Bennu, OLA became indispensable for terrain modeling, hazard assessment, analyses by the Altimetry Working Group (ALTWG), and development of Natural Feature Tracking, making it a critical contributor to successful sample acquisition. This experience demonstrated that preserving high-value international contributions often requires flexible governance, proactive communication, and collaborative problem-solving rather than rigid adherence to initial implementation plans.

These experiences demonstrated that mission success depends not only on technical excellence but also on ensuring that organizational authority, contractual relationships, acquisition strategies, and international partnerships remain aligned with the mission's management philosophy and evolving technical needs.

## Lessons Learned

- **Align organizational authority and contractual relationships with the intended mission management structure**. When a Principal Investigator is responsible for mission success, authority, communication pathways, and access to implementing organizations should enable timely technical and programmatic decision-making rather than create unnecessary organizational barriers.
- **Match acquisition strategy to the technical maturity, organizational experience, and mission criticality of the systems being procured**. Higher-risk contractual approaches require commensurate investments in management oversight, technical support, systems engineering, contingency planning, and independent backup capabilities.
- **Manage international partnerships proactively** throughout the mission lifecycle. Funding uncertainties, schedule changes, and evolving technical requirements should be addressed through continuous communication, collaborative governance, and established decision processes that preserve mission-critical capabilities whenever practical.

- **Recognize that authority, contracts, and partnerships function as an integrated management system**. Organizational authority, contractual relationships, acquisition strategies, and institutional partnerships should reinforce one another, enabling rapid issue resolution while preserving mission-critical capabilities as technical and programmatic conditions evolve.

---

# 2. Cost, Schedule, and Risk Management

## 2.1 Integrated Risk Management and Early Descope Planning

**Key Theme:** Major technical, cost, schedule, and programmatic risks should be identified early, with executable contingency plans and clearly defined decision criteria that preserve mission success while protecting mission-critical capabilities.

### Background

Risk management was fully integrated into the OSIRIS-REx management approach from the earliest phases of formulation. Technical, cost, schedule, and programmatic risks were continuously evaluated throughout development using a structured risk management process that linked each significant risk to documented mitigation strategies, contingency plans, reserve posture, and objective decision criteria.

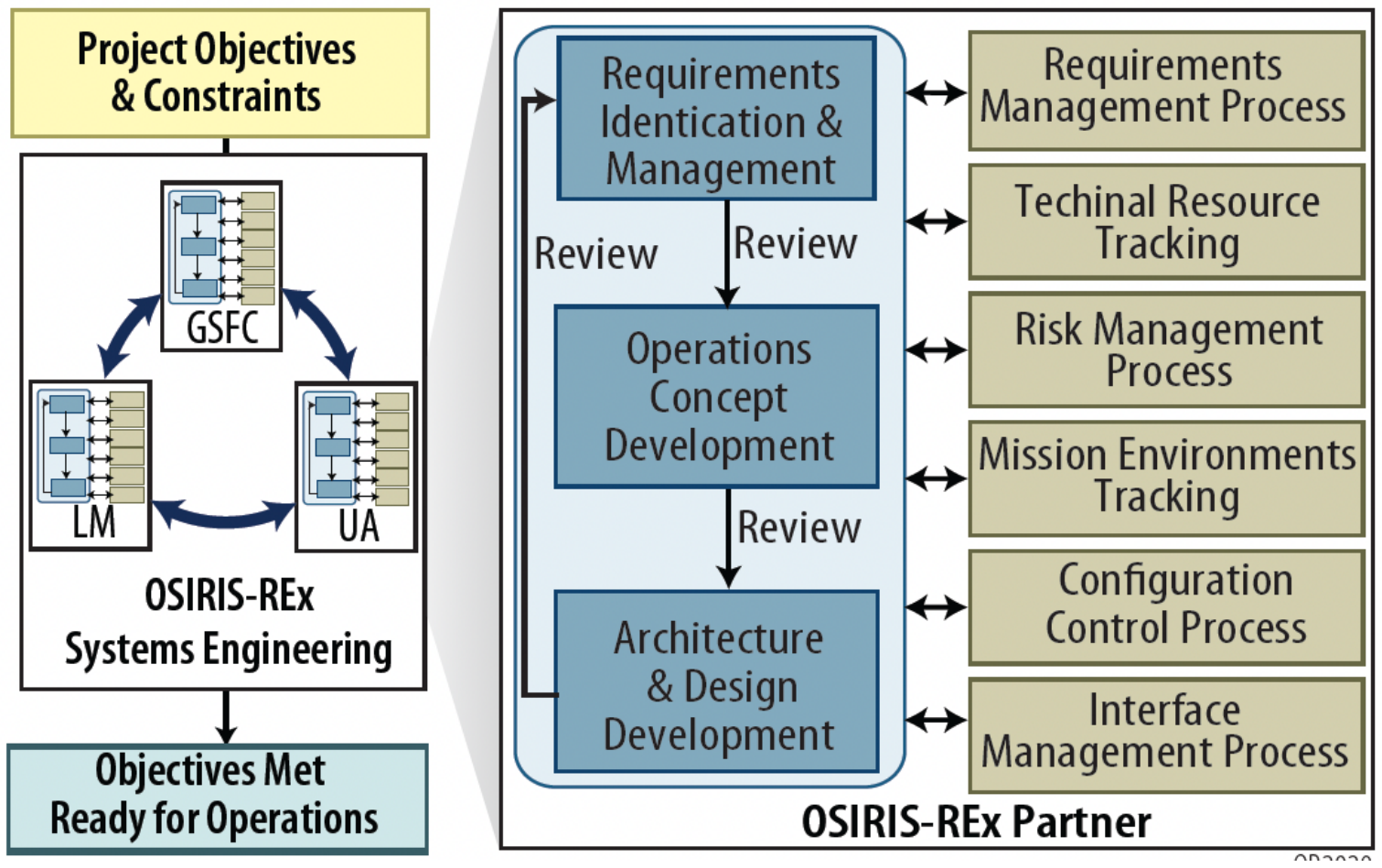

**Figure 2**. The integration of the OSIRIS-REx partners' System Engineering processes minimized costs. Credit: OSIRIS-REx Concept Study Report.

A defining feature of this process was the development of an explicit descope strategy before it was needed. Rather than treating descopes as reactive responses to budget pressure, the PI Management Team identified candidate descopes early, quantified their expected cost and schedule savings, assessed their impact on scientific return and mission objectives, and established decision trigger points tied to reserve posture and technical risk. This disciplined planning ensured that difficult decisions could be made deliberately rather than under crisis conditions.

The value of this approach was demonstrated when one payload instrument experienced significant requirements growth accompanied by escalating cost and schedule impacts. Although removing a scientific instrument was difficult and not universally supported, detailed technical and scientific analyses demonstrated that all Level-1 mission objectives could still be achieved without the instrument. Because objective decision criteria and executable descope plans had already been established, the project acted early enough to preserve cost and schedule reserves while maintaining the mission's primary objectives.

OSIRIS-REx also demonstrated that effective risk management requires recognizing when mission-critical capabilities should be protected rather than descoped. As discussed in Section 1.4 Organizational Authority, Contract Alignment, and International Partnerships the Canadian Space Agency's OSIRIS-REx Laser Altimeter (OLA) initially represented a significant programmatic risk because of uncertainties in international funding and schedule coordination. Rather than accepting an early descope, the PI Management Team worked collaboratively with NASA, the Canadian Space Agency, and MDA to develop flexible implementation strategies that preserved the instrument within the project baseline.

These complementary decisions illustrate that successful risk management is not simply about reducing scope. It requires disciplined, evidence-based decision making that preserves the highest overall probability of mission success while maintaining the flexibility to protect capabilities whose true value may emerge only as the mission evolves.

## Lessons Learned

- **Integrate risk management into every phase** of the mission lifecycle. Identify major technical, cost, schedule, and programmatic risks early, and maintain documented mitigation strategies, contingency plans, and decision criteria throughout project execution.
- **Develop executable descope plans before they are needed**. Candidate descopes should include quantified assessments of cost savings, schedule impact, technical performance, scientific return, and clearly defined decision trigger points linked to project reserve posture.
- **Execute difficult decisions early** when objective decision criteria indicate that mission success is best served by protecting overall project health. Delaying necessary descopes often increases cost, schedule, and technical risk while reducing available management options.
- **Preserve mission-critical capabilities** whose long-term operational value outweighs near-term programmatic pressures. Flexible management, strong partnerships, disciplined systems engineering, and collaborative problem solving can prevent premature descopes of capabilities that later become essential to mission success.

---

## 2.2 Integrated Cost Planning and Estimating

**Key Theme:** Inclusive cost planning, disciplined Basis of Estimate (BOE) development, and integrated cost management produce credible mission budgets and improve cost accountability throughout the mission lifecycle.

### Background

From the earliest Concept Study Report (CSR) through mission operations, OSIRIS-REx employed an integrated cost planning process centered on recurring Cost Summits. Rather than developing budgets independently within individual organizations, all major mission elements—including the University of Arizona, NASA Goddard Space Flight Center, Lockheed Martin, and instrument teams—participated in structured cost planning using a common set of planning assumptions and mission products.

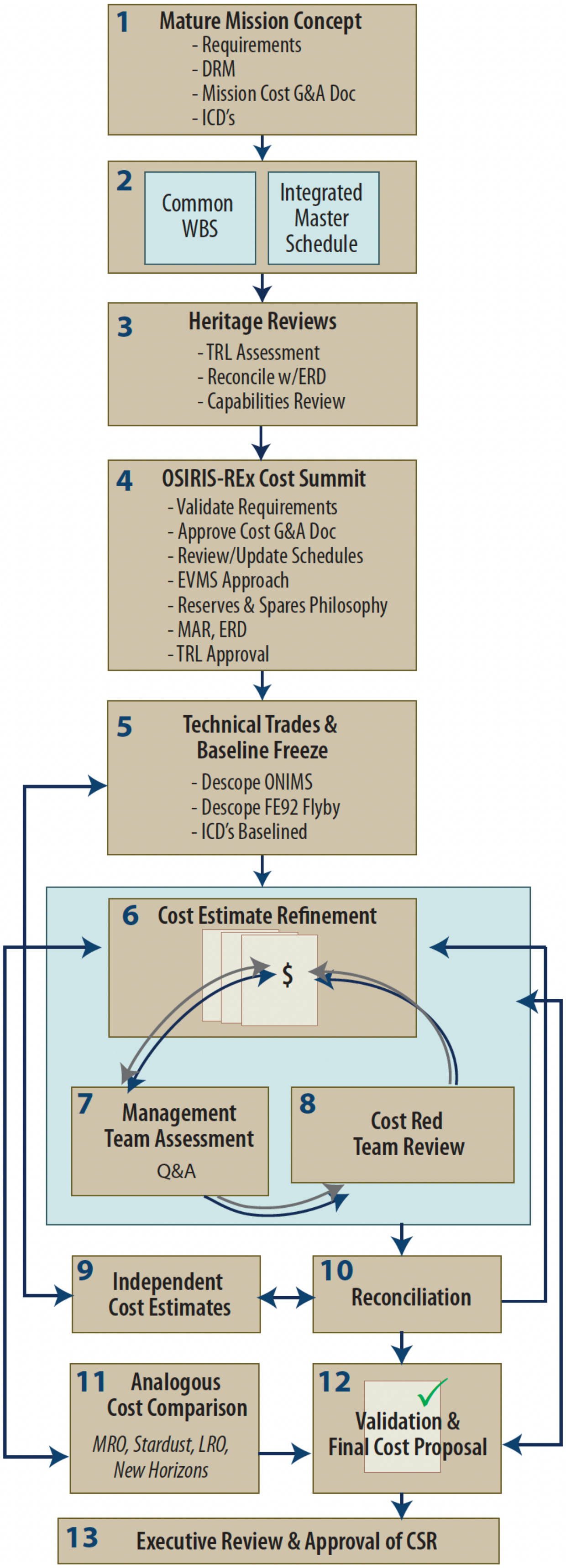

**Figure 3**. OSIRIS-REx followed a rigorous and interactive costing process that resulted in a high-fidelity cost estimate. Credit: OSIRIS-REx Concept Study Report.

Each Cost Summit was built upon a shared technical foundation that included the Design Reference Mission (DRM), Work Breakdown Structure (WBS), Statements of Work (SOWs), Integrated Master Schedule (IMS), systems engineering products, and common Guidelines and Assumptions. This ensured that all organizations estimated cost from the same technical baseline rather than from independent interpretations of project scope.

The process emphasized development of detailed Basis of Estimate (BOE) documentation for every major WBS element. BOEs documented the assumptions, staffing profiles, deliverables, schedules, risks, and estimating methodologies supporting each budget request. These estimates were reviewed collaboratively during Cost Summits, allowing assumptions to be challenged, inconsistencies identified, and interfaces reconciled across organizational boundaries.

OSIRIS-REx also adopted a hybrid cost estimating methodology that combined three complementary approaches. Detailed grassroots BOEs formed the primary estimate for each WBS element. Parametric cost models were used to independently validate those estimates, and carefully selected analog missions provided additional context for evaluating mission-specific assumptions. No single estimating method was considered sufficient; confidence was achieved through comparison, reconciliation, and engineering judgment across multiple independent approaches. This methodology proved particularly valuable for estimating the highly unique Phase E science operations, where few truly comparable missions existed.

Between formal Cost Summits, a standing Cost Working Group, led by the Project Planning and Control Office, met regularly with WBS managers to monitor cost performance, update BOEs, assess emerging risks, and maintain alignment between evolving technical scope and available resources. This continuous engagement transformed cost estimation from a one-time proposal activity into an ongoing project management discipline.

The result was a transparent, technically grounded, and broadly supported cost baseline that evolved with the mission while maintaining consistency among science objectives, technical scope, schedule, and available budget.

## Lessons Learned

- **Conduct recurring Cost Summits** that bring together all major project organizations using a common technical baseline, shared planning assumptions, and standardized estimating practices. Collaborative cost planning produces more credible budgets than independently developed organizational estimates.
- **Develop and maintain detailed Basis of Estimate documentation** for every major Work Breakdown Structure element. BOEs should explicitly document assumptions, staffing, deliverables, schedules, risks, and estimating methodologies so that cost estimates remain traceable throughout the project lifecycle.
- **Employ multiple, complementary estimating methodologies**—including grassroots estimates, parametric models, and carefully selected analog missions—to improve confidence, identify inconsistencies, and strengthen engineering judgment. No single estimating method is sufficient for complex flagship missions.
- **Maintain a standing Cost Working Group** between formal planning milestones to update BOEs, monitor emerging risks, reconcile evolving technical scope, and preserve alignment among cost, schedule, and systems engineering.
- **Treat cost estimation as an integrated systems engineering activity** rather than an independent financial exercise. Transparent collaboration among project management, engineering, science, and planning organizations strengthens accountability, improves budget realism, and enables more informed programmatic decisions.

## 2.3 Strategic Reserve and Margin Management

**Key Theme:** Financial reserves and schedule margin should be established early and managed strategically throughout the mission lifecycle to reduce risk, preserve flexibility, and enable timely decision-making.

### Background

OSIRIS-REx recognized from formulation that reserve management is not simply a budgeting exercise but a fundamental element of mission risk management. Financial reserves and schedule margin were deliberately established early to provide the project with the flexibility needed to respond to technical challenges, evolving mission requirements, and unforeseen external events.

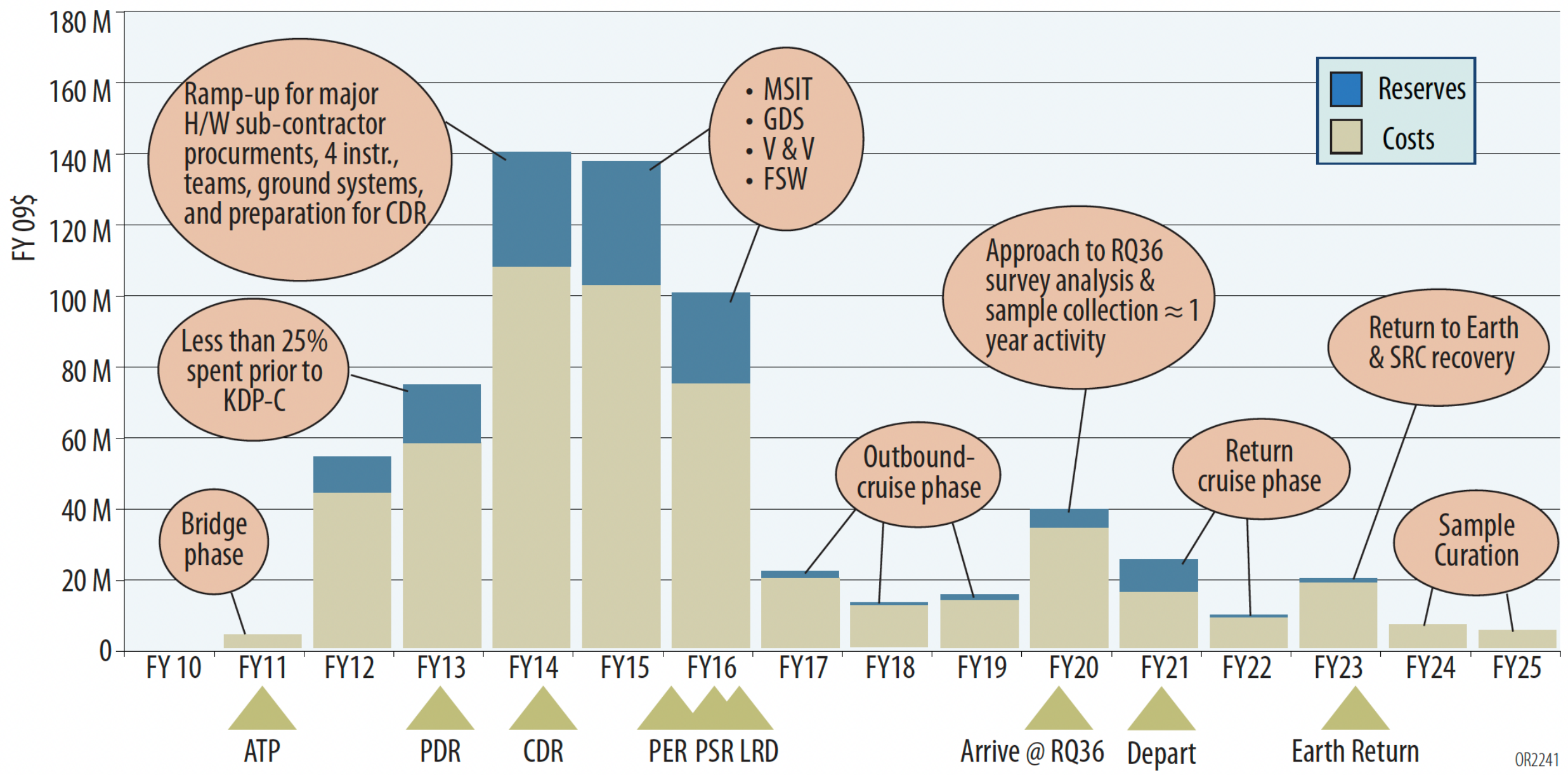

**Figure 4**. Cost Profile by FY fully supported the successful delivery of the OSIRIS-REx mission. Footnote: Totals are FY09$. Credit: OSIRIS-REx Concept Study Report.

Rather than distributing available resources uniformly across the project, funding was intentionally front-loaded to ensure that critical development activities, risk-reduction efforts, and technical problem resolution could proceed while corrective action remained practical and affordable. This philosophy recognized that reserves create the greatest value when invested early enough to reduce risk rather than merely recover from problems after they occur.

Reserve management remained closely coupled to the project's evolving technical risk. As uncertainties were retired and confidence increased, reserves were released deliberately to support emerging priorities while maintaining sufficient protection against future unknowns. A defining example was the early decision to fund development of Natural Feature Tracking (NFT) as an independent backup to the LIDAR navigation system (Lauretta et al., 2017; Lorenz et al., 2017). Although NFT represented a significant investment during development, releasing reserves early allowed the capability to mature before arrival at Bennu. When the asteroid's unexpectedly rugged surface rendered the original, LIDAR-based sampling strategy inadequate, NFT became the enabling technology for safe sample acquisition and ultimately a primary contributor to mission success (Getzandanner et al., 2022; Lauretta et al., 2021, 2022; Norman et al., 2022).

The value of maintaining  financial reserves and schedule margin was demonstrated repeatedly throughout the mission. Early reserve allocation enabled the project to respond effectively to technical issues, hardware rework, vendor challenges, federal government shutdowns, and other unforeseen events without compromising mission objectives. The combination of disciplined reserve management, conservative schedule planning, and integrated cost governance provided the resilience needed to adapt to changing circumstances while maintaining overall project performance.

## Lessons Learned

- **Establish financial reserves and schedule margin early** in the project lifecycle so they can be used proactively to reduce technical, cost, and schedule risk rather than reactively after problems have escalated.
- **Manage reserve allocation as a risk-reduction strategy** rather than a budget contingency. Reserve release should be driven by evolving technical risk, project readiness, and opportunities to reduce future uncertainty—not by calendar milestones.
- **Invest reserves early enough to influence mission outcomes**. Funding critical risk-reduction activities during development often provides substantially greater benefit than preserving reserves until late in the project, when management options have become limited.
- **Protect schedule margin as a strategic project resource**. Adequate schedule flexibility enables informed engineering decisions, accommodates unforeseen technical developments, and reduces pressure to accept unnecessary mission risk.

- **Develop a unified risk management framework** that integrates financial reserves and schedule margin. Together, they provide the resilience needed to adapt to changing technical, operational, and programmatic conditions while preserving mission success.

---

## 2.4 Schedule Realism

**Key Theme:** Development, operations, curation, and sample analysis all require realistic schedules with sufficient margin to accommodate discovery, uncertainty, and evolving mission requirements.

### Background

OSIRIS-REx adopted a conservative scheduling philosophy that recognized that exploration missions cannot be executed successfully using optimistic timelines. The baseline Design Reference Mission (DRM), developed during formulation, intentionally incorporated sufficient schedule margin to support comprehensive asteroid characterization before committing to sample collection. Rather than treating schedule as a fixed constraint, the DRM served as a living systems engineering document that evolved as knowledge of Bennu increased and operational assumptions changed.

This philosophy proved essential after arrival at Bennu. The asteroid's unexpectedly rugged surface required additional reconnaissance, detailed hazard mapping, development of Natural Feature Tracking (NFT), refinement of sample-site selection, and modifications to the operational timeline (Burns & Lauretta, 2025; Getzandanner et al., 2022; Norman et al., 2022). Because adequate schedule margin had been incorporated into the mission design, these activities could be completed without compromising mission objectives or exceeding the approved project budget. The project also successfully accommodated disruptions associated with the COVID-19 pandemic while maintaining readiness for the sample collection campaign (Burns & Lauretta, 2025; Getzandanner et al., 2022). Rather than delaying the mission, schedule margin enabled better scientific understanding and more informed operational decisions.

OSIRIS-REx also demonstrated that sample-return missions encompass three distinct scientific campaigns: pre-launch astronomical characterization, in situ exploration at the target body, and laboratory investigation of the returned samples. While the development and flight phases incorporated appropriate schedule margin, the original mission plan allocated only two years for initial sample analysis. Once the returned material was available, however, it became clear that the scientific complexity, diversity, and long-term value of the collection required substantially more time. Unlike the earlier mission phases, sample analysis had not been planned as a fully integrated mission phase with sufficient schedule flexibility and reserves to accommodate scientific discovery. Experience from OSIRIS-REx has since influenced NASA policy by recognizing sample analysis as an integral component of sample-return missions rather than a post-mission activity. (See Section 5.1 Sample Analysis Planning for details.)

These experiences demonstrate that exploration missions require schedule realism not simply because technical problems occur, but because scientific discovery changes what missions know, what they prioritize, and how they operate. Schedules should therefore be designed to accommodate learning rather than assume that mission knowledge remains fixed after formulation.

### Lessons Learned

- **Resist schedule optimism** and incorporate realistic schedule margin from formulation through mission completion. Schedule flexibility enables projects to respond effectively to evolving scientific knowledge, operational realities, and unforeseen events while preserving mission objectives.
- **Maintain the Design Reference Mission as a living document** throughout the project lifecycle. As scientific understanding and operational constraints evolve, mission plans should be updated while maintaining alignment among technical requirements, available resources, cost, schedule, and acceptable risk.
- **Recognize that exploration missions are driven by discovery.** Schedule margin should provide sufficient time to understand new information before making irreversible engineering or operational decisions.
- **Plan sample curation and scientific analysis with the same management rigor** applied to spacecraft development and flight operations. Schedule margin, financial reserves, staffing, and project management should extend through the scientific investigation phase to accommodate the inevitable discoveries arising from returned extraterrestrial materials.

---

# 3. Systems Engineering and Technical Management

## 3.1 Interface Management

**Key Theme:** Complex, distributed missions require rigorous interface definition, ownership, communication, and integration planning throughout the mission lifecycle.

### Background

OSIRIS-REx was executed by a geographically distributed team that included the University of Arizona, NASA Goddard Space Flight Center, Lockheed Martin, KinetX, instrument teams, and numerous science and engineering organizations. Mission success depended not only on the performance of individual systems but also on the effectiveness of the interfaces connecting organizations, hardware, software, operational products, and decision-making processes.

Recognizing this complexity, the project invested heavily in interface management from the earliest phases of development. Critical technical interfaces, data products, operational workflows, and communication pathways were identified and documented through systems engineering processes, while recurring cross-functional meetings and integrated planning activities maintained alignment as the mission evolved. A history of collaboration among several of the partner organizations further reduced integration risk by providing established data flows, operational procedures, and trusted working relationships.

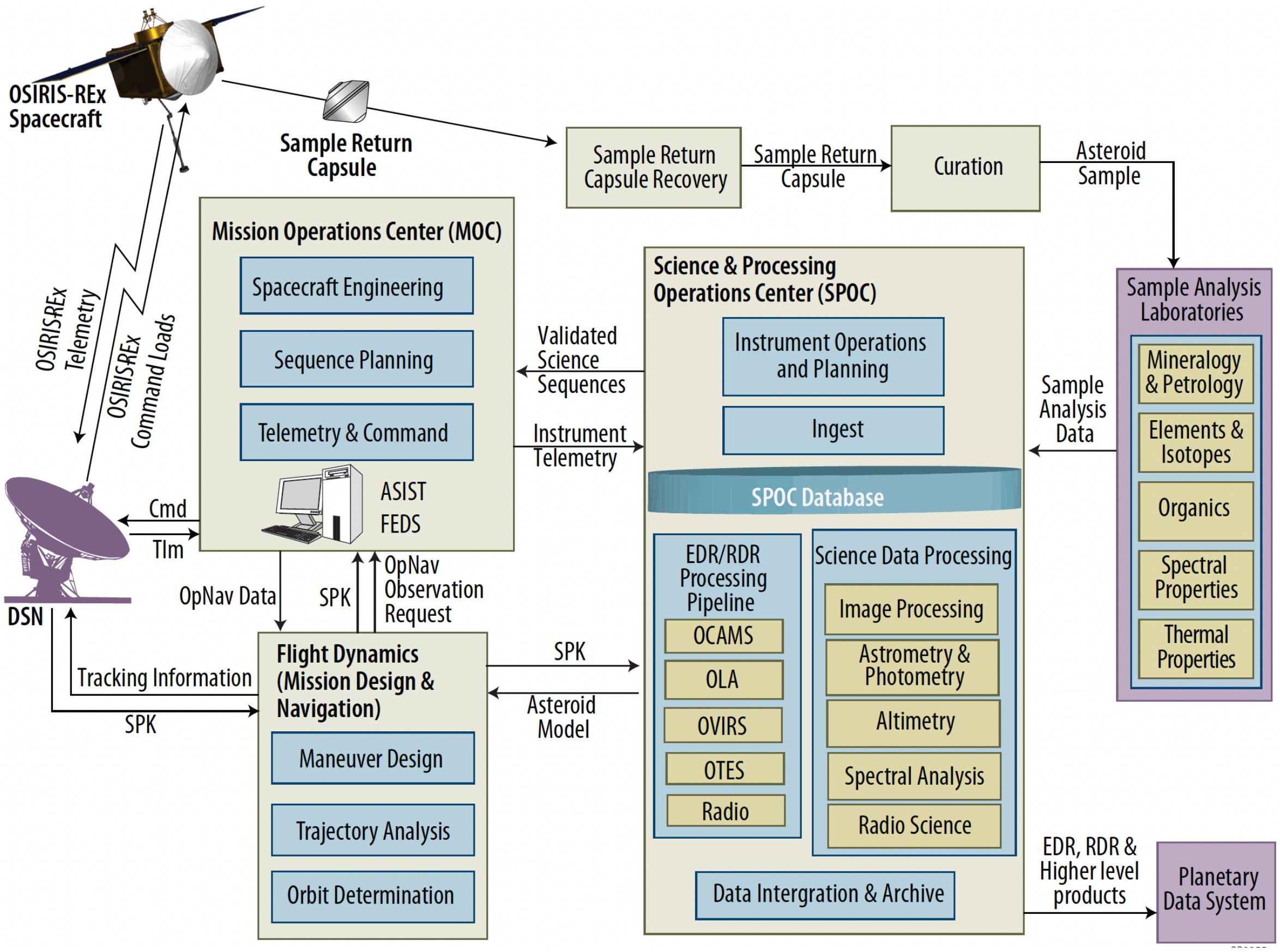

**Figure 5**. Mission Operations Architecture. Ground Segment functions, interfaces, and data flows applied heritage to the OSIRIS-REx mission design. Credit: OSIRIS-REx CSR.

The importance of rigorous interface management became particularly evident during Phase E operations. Flight dynamics, optical navigation, science operations, spacecraft engineering, and mission planning operated as a tightly coupled system in which delays or misunderstandings at one interface could propagate rapidly throughout the mission. Although these interfaces functioned successfully overall, the project found that several activities—notably flight dynamics and optical navigation—had been underestimated in both scope and staffing because of the complexity of the interfaces they supported. Interface management therefore required not only technical definition but also realistic allocation of schedule, staffing, and management attention.

The development of Natural Feature Tracking (NFT) further demonstrated the importance of interface engineering. The Altimetry Working Group (ALTWG) became responsible for integrating laser altimetry, imaging, digital terrain (shape) models, navigation products, and flight software into a unified NFT catalog. Initially, the working group's responsibilities, authority, staffing, and communication pathways were not fully defined, leading to coordination challenges and planning gaps between ALTWG, the Flight System, and the Operations Ground System. Once organizational responsibilities were clarified and interface ownership established, collaboration improved rapidly and NFT development proceeded efficiently. The experience demonstrated that interface management encompasses organizational structure and decision authority as much as technical compatibility.

Throughout the mission, successful interface management depended on disciplined systems engineering, clearly defined responsibilities, continuous communication, and proactive integration planning. These practices enabled the project to coordinate among numerous organizations while maintaining the agility needed to respond to evolving technical and operational requirements.

## Lessons Learned

- **Define interface ownership, organizational authority, communication pathways, and decision responsibilities early**, particularly for cross-functional activities spanning multiple institutions or technical disciplines.
- **Treat interface management as a continuous systems engineering discipline** rather than a one-time development activity. Technical, organizational, operational, and software interfaces should be identified, documented, and actively managed throughout the mission lifecycle.
- **Scope interface-intensive organizations realistically**. Activities such as flight dynamics, optical navigation, and integrated mission planning may require substantially greater staffing and schedule resources than initially estimated because they depend on continuous coordination across multiple organizations.
- **Establish cross-disciplinary working groups** with clearly defined authority whenever mission success depends on integrating products from multiple organizations. Working groups should possess sufficient resources, decision authority, and management visibility to resolve interface issues before they affect mission execution.
- **Recognize that successful interface management depends equally on technical integration, organizational alignment, and effective communication**. Well-defined interfaces reduce ambiguity, improve execution efficiency, and enable rapid response to changing mission conditions.

---

## 3.2 Designing for Adaptability

**Key Theme:** Mission architectures should be designed to accommodate unexpected environmental conditions, changing requirements, and evolving operational knowledge.

### Background

OSIRIS-REx demonstrated that flexibility is not a contingency feature—it is a fundamental systems engineering requirement for exploration missions. Despite extensive pre-launch characterization and careful mission planning, many of the assumptions that informed spacecraft design and operations changed after arrival at Bennu. The mission succeeded because adaptability had been deliberately incorporated into the spacecraft design, flight software, operational concepts, and systems engineering processes from the earliest phases of development.

For example, flexible thermal design accommodated more stringent sample preservation requirements that emerged during development without requiring major redesign. Spacecraft interfaces developed during Phase C included provisions ("hooks") that later enabled integration of Natural Feature Tracking (NFT), allowing the navigation system to evolve in response to operational needs.

Following arrival at Bennu, the mission encountered a series of unexpected conditions that fundamentally altered operational planning. The asteroid's surface proved substantially more rugged than anticipated, requiring complete revision of the sample-site selection process, significantly more accurate navigation, and the development of autonomous hazard avoidance capabilities (Burns & Lauretta, 2025; Lauretta et al., 2019a, 2021). The unexpected discovery of particle ejection events introduced a sudden potential operational hazard while simultaneously creating new scientific opportunities (Lauretta et al., 2019b). Rather than forcing the mission to conform to its original assumptions, the project adapted its operational concepts, engineering solutions, and scientific priorities to reflect the reality of Bennu's environment.

The success of these adaptations was made possible because flexibility had been incorporated into the mission architecture long before specific changes were known. Systems engineering emphasized margin, modularity, and operational adaptability, allowing new capabilities to be introduced without compromising spacecraft safety or mission objectives. The experience demonstrated that exploration missions should be designed to accommodate discovery

### Lessons Learned

- **Design mission architectures to accommodate evolving knowledge** rather than fixed assumptions. Exploration missions should anticipate that target environments, operational requirements, and scientific priorities will differ from pre-launch paradigms.
- **Incorporate flexibility** into hardware, software, operational concepts, and systems engineering products from the earliest phases of development. Adaptability is most effective when designed into the system rather than added after problems emerge. Systems designed for flexibility are more resilient, better able to exploit new discoveries, and more likely to achieve mission objectives under uncertain conditions.
- **Provide architectural provisions for future capabilities** whenever practical. Early investment in interface flexibility, modular software, and expandable operational concepts enables rapid response to unforeseen mission challenges.
- **Maintain sufficient engineering margin** to accommodate changing environmental conditions, revised technical requirements, and emerging scientific opportunities without requiring fundamental redesign.

---

## 3.3 Ground Systems Development

**Key Theme:** Ground systems require the same systems engineering rigor, staffing, management attention, and requirements discipline as flight systems.

### Background

OSIRIS-REx demonstrated that ground systems are mission-critical engineering systems whose development must proceed in parallel with spacecraft and instrument development. Mission success depended on an integrated ground system capable of supporting science planning, mission operations, navigation, data processing, observation design, and decision-making throughout the mission lifecycle.

From the outset, the project emphasized an integrated systems engineering culture in which scientific objectives, flight systems, ground systems, and operational products were developed using common requirements, shared technical products, and transparent communication. Frequent cross-functional coordination ensured that evolving science priorities, navigation products, and operational requirements remained aligned across all participating organizations. This systems engineering approach strengthened transparency, improved technical understanding, and enabled the project to reprioritize observations and operational activities efficiently as new information became available.

Despite this strong systems engineering foundation, the project initially underestimated the complexity of ground system development. The original implementation of the Science Processing and Operations Center (SPOC) did not evolve in step with the rapidly changing mission architecture, resulting in significant redesign, software redevelopment, and organizational restructuring. Experience demonstrated that ground systems must evolve continuously alongside mission design rather than being treated as downstream support infrastructure.

A similar lesson emerged during science planning and operations development. The staffing, systems engineering effort, and operational planning required to support observation design, navigation integration, and mission execution proved substantially greater than originally anticipated. Because Bennu operations permitted little opportunity to repeat missed observations, mission success depended upon rigorous planning, comprehensive validation, and continuous refinement of operational products. Ground systems therefore became active components of mission execution rather than passive data-processing tools.

Overall, OSIRIS-REx demonstrated that successful ground systems require the same disciplined systems engineering, realistic staffing, management visibility, configuration control, and lifecycle planning applied to spacecraft development. Their complexity grows throughout the mission and must be managed accordingly.

## Lessons Learned

- **Treat ground systems as mission-critical** engineering systems whose development, verification, staffing, and management receive the same rigor and visibility as spacecraft and instrument development.
- **Integrate ground system development with evolving mission architecture** from the earliest phases of the project. Requirements, interfaces, operational concepts, and software capabilities should evolve together rather than independently.
- **Resource ground systems realistically**. Science planning, operations planning, navigation support, software development, and operational validation frequently require substantially greater staffing and systems engineering effort than initially estimated. Planning, management attention, and lifecycle support should anticipate continuous refinement rather than assuming development is complete before launch.

- **Maintain continuous communication** among science, engineering, navigation, flight operations, and ground system organizations. Transparent requirements management and shared technical products enable rapid adaptation as mission knowledge evolves.

---

## 3.4 Functional Redundancy and Capability Flexibility

**Key Theme:** Understanding overlapping instrument capabilities creates operational resilience by enabling systems to assume new roles when primary capabilities are degraded or lost.

### Background

OSIRIS-REx demonstrated that mission resilience depends not only on explicit hardware redundancy but also on understanding the functional overlap among spacecraft subsystems and scientific instruments. As the mission evolved, several instruments successfully assumed responsibilities beyond their original design, allowing critical operations to continue despite hardware limitations and changing mission requirements.

One example emerged during development when it became apparent that the spacecraft's Navigation Cameras (NavCams), a critical element of the optical navigation system, had not been designed for the radiation environment expected during the mission. To mitigate this vulnerability, the project promoted MapCam, originally developed as part of the OCAMS science instrument suite (Rizk et al., 2018), to serve as the primary optical navigation camera. This change required modifications to MapCam's lifetime requirements, additional qualification testing, and expanded operational support during Phase E, but ultimately provided a robust and reliable navigation capability.

A second example occurred during asteroid operations when the Low-Energy Laser Transmitter (LELT) within the OSIRIS-REx Laser Altimeter (OLA) failed during the Reconnaissance phase, when data were being collected to enable Natural Feature Tracking (NFT). Rather than compromising development of NFT, the project repurposed the OCAMS PolyCam to provide the necessary optical focus measurements. This adaptation was possible because the project had developed a detailed understanding of the overlapping capabilities of the payload, making it possible to identify an alternative technical solution before the loss became mission-limiting (Lauretta et al., 2021).

These experiences demonstrated that functional redundancy extends beyond duplicate hardware. A comprehensive understanding of instrument capabilities, supported by cross-disciplinary systems engineering and operational creativity, enables missions to recover from unexpected failures while preserving critical functions.

### Lessons Learned

- **Design mission-critical systems with appropriate environmental robustness** for their intended operational roles. Components supporting navigation, guidance, or other essential spacecraft functions should be qualified for the environments they will encounter throughout the mission.
- **Identify overlapping capabilities across spacecraft subsystems and scientific instruments** early in development. Functional redundancy can often provide effective operational resilience even when explicit hardware redundancy is unavailable.
- **Maintain systems-level understanding of payload capabilities** so that instruments and subsystems can assume new operational roles as mission needs evolve. Cross-disciplinary collaboration among instrument teams, flight systems, navigation, and science operations can reveal innovative solutions to unexpected technical challenges.
- **Treat functional redundancy as an integral component of mission resilience**. Designing for flexibility in how capabilities are used can significantly increase robustness while minimizing additional cost and complexity.

---

## 3.5 Heritage Hardware and Heritage Designs

**Key Theme:** Heritage hardware and designs must be independently verified and validated. Documentation inheritance is not a substitute for preserving engineering knowledge, and mission-critical systems should incorporate layered redundancy.

### Background

OSIRIS-REx deliberately leveraged heritage hardware and proven designs to reduce technical risk, cost, and development time. Although heritage systems provide substantial advantages, the mission demonstrated that successful reuse requires more than inheriting design documentation. Projects must verify that they understand the complete as-built and as-flown implementation, including undocumented modifications, engineering rationale, and operational experience that may not be fully captured in formal records.

This lesson became evident during investigation of the Sample Return Capsule (SRC) parachute deployment anomaly. The capsule architecture was derived directly from that of the Stardust mission (Williams et al., 2025), and the project assumed that the interface between the electronics assembly and the parachute release system was fully understood. Because the Stardust capsule had landed safely without incident, portions of the original end-to-end verification testing were not repeated for OSIRIS-REx. Unknown to the project, late changes made during the Stardust development had corrected an error in the wiring documentation after engineering flight testing revealed crossed deployment circuits. Those updates were not fully propagated into the documentation inherited by OSIRIS-REx, and much of the engineering expertise associated with those changes had been lost during the 12 years separating the two missions. As a result, the inherited documentation did not accurately represent the final flight configuration (NASA, 2023; Williams et al., 2025).

Fortunately, the Sample Return Capsule had been designed with independent redundancy in its parachute deployment architecture. Main parachute deployment could be initiated either by a timer or by an independent pressure-based trigger. When the drogue parachute failed to deploy as expected, the capsule descended to the nominal main parachute deployment altitude before the timer expired. The pressure trigger activated the main parachute automatically, allowing the capsule to land safely despite off-nominal entry conditions (Lauretta, Connolly, et al., 2024; NASA, 2023; Williams et al., 2025). Although post-flight inspection showed that the parachute experienced significantly higher loads than anticipated, the redundant deployment logic and structural design margins successfully protected the vehicle and its scientific payload.

Together, these experiences demonstrate that heritage reduces risk only when accompanied by rigorous verification and preservation of engineering knowledge, and critical mission functions should be protected by independent layers of redundancy capable of tolerating unforeseen failures.

## Lessons Learned

- **Verify heritage hardware and designs** through independent end-to-end testing rather than relying solely on inherited documentation or previous mission success.
- **Preserve engineering knowledge** as an essential component of heritage. Projects should actively record design rationale, implementation details, test history, and lessons learned, such that future missions can acquire a complete basis even when significant time has elapsed.
- **Reconstruct the complete as-flown configuration** of heritage systems whenever possible. Documentation, drawings, and configuration records alone may not fully capture modifications made during previous development or test campaigns.

- **Design mission-critical systems with independent, layered redundancy** wherever practical. Redundant sensing, triggering, or control paths provide resilience against unforeseen failures and reduce the likelihood that a single fault will jeopardize mission success.
- **Validate heritage assumptions continuously**. Previous flight success demonstrates that a system worked under one set of conditions; it does not eliminate the need to verify that the inherited design remains fully understood and appropriate for a new mission.

---

# 4. Mission Operations

## 4.1 Trust, Transparency, and Communication

**Key Theme:** Transparent communication and mutual trust are foundational elements of operational resilience.

### Background

OSIRIS-REx demonstrated that operational resilience depends on more than procedures, tools, and personnel. During high-stakes mission operations, teams must be able to exchange information rapidly, trust one another's expertise, and make decisions under time pressure with incomplete information. This capability was built over years of direct engagement, shared problem solving, and cross-functional collaboration.

The importance of this foundation was demonstrated during several unexpected events. During the first orbit insertion, a loss of network connectivity required the team to rapidly pivot to an alternate pathway using JPL systems to uplink critical command files. The response succeeded because roles were clear, decision authority was understood, and the team trusted one another's technical judgment.

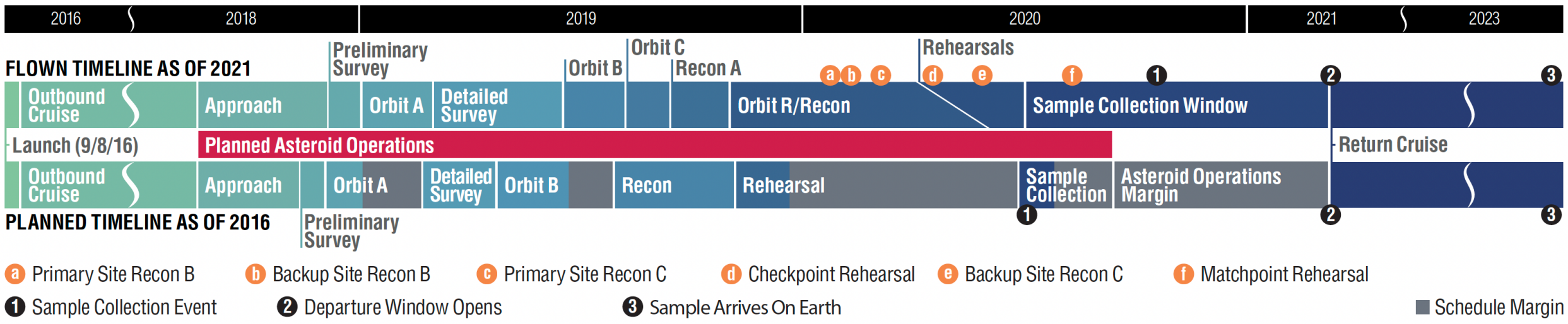

**Figure 6**. Comparison of the as-flown (top row) and planned (bottom row) mission timelines, beginning with launch in September 2016 and extending to sample return in September 2023. From: Lauretta et al. (2021), In *Sample return missions* (pp. 163-194). Elsevier. Credit: University of Arizona.

The onset of the COVID-19 pandemic in early 2020 provided a second major test. While preparing for the Touch-and-Go (TAG) sample collection maneuver , the team transitioned to distributed operations with minimal disruption. This success reflected the strong culture of trust and collaboration that had been established before the crisis, allowing personnel to adapt procedures, maintain communication, and continue mission-critical work under unprecedented constraints.

OSIRIS-REx also showed that trust can be weakened when tools, assumptions, or analytical methods are opaque to partner organizations. Tensions between the science planning, mission systems assurance, and flight dynamics teams were amplified when key tools or decision processes were not fully visible or understood across organizational boundaries. Increased face-to-face engagement, co-location, shared technical explanations, and the introduction of tools such as the Activity Inspector helped bridge these gaps, improving mutual understanding and reducing friction.

### Lessons Learned

- **Build trust early**. Relationships cultivated during formulation, development, and rehearsals become essential assets of operational resilience. Technical success under pressure often depends on mutual trust, personal engagement, and shared situational awareness (Getzandanner et al., 2022).
- **Ensure visibility and comprehensibility** of operational tools, analytical methods, and decision criteria to affected organizations. Transparency reduces friction, improves confidence, and strengthens cross-functional decision-making.
- **Maintain clear roles, responsibilities, and decision authority** during both nominal and off-nominal operations. Teams can respond rapidly to unexpected events when they understand who is empowered to act.
- **Resolve organizational tension before it becomes operational risk by means of** direct engagement, co-location, and shared technical walkthroughs.

---

## 4.2 Co-Location and Cross-Functional Integration

**Key Theme:** Physical co-location and integrated cross-functional teams enable rapid decision-making, effective communication, and agile response during complex mission operations.

## Background

Operations at Bennu proved substantially more complex than anticipated, requiring continuous replanning, rapid technical assessments, and real-time coordination across multiple organizations. Scientific observations, spacecraft operations, flight dynamics, optical navigation, systems engineering, and mission planning became tightly coupled activities in which decisions often had to be made on timescales that left little opportunity for formal coordination through traditional organizational channels.

A major contributor to mission success was the deliberate co-location of science planning, science working Group leadership, Flight Dynamics, spacecraft engineering, and mission operations personnel during critical operational periods. Physical proximity enabled immediate communication, rapid exchange of technical information, and collaborative problem solving without the delays associated with geographically distributed decision-making. Questions that might otherwise have required multiple meetings or lengthy coordination could often be resolved within minutes through direct interaction among the appropriate technical experts.

Equally important, as discussed in Section 4.1 Trust, Transparency, and Communication, the effectiveness of co-location depended upon trusted professional relationships that had been established throughout the development phase. Years of collaboration created a shared understanding of technical objectives, operational constraints, and organizational responsibilities. This foundation enabled teams to challenge assumptions openly, evaluate alternative solutions quickly, and adapt operational plans while maintaining mission safety and scientific priorities.

The combination of integrated operations, direct communication, and mature working relationships allowed OSIRIS-REx to respond effectively to the unexpected complexity of Bennu's environment while maintaining operational efficiency throughout proximity operations and sample acquisition.

## Lessons Learned

- **Co-locate core operational organizations** whenever practical during complex mission phases. Physical proximity significantly improves communication efficiency, decision speed, and cross-disciplinary problem solving.
- **Integrate mission elements into a unified operational team** rather than relying on sequential organizational handoffs. Effective mission operations depend as much on organizational integration as on technical excellence; well-integrated teams are better able to respond to uncertainty while maintaining mission safety and scientific return.

- **Invest in building working relationships before critical operations begin**. Mutual trust and shared technical understanding are essential enablers of rapid, high-quality decision-making under operational pressure.
- **Empower cross-functional teams to resolve issues** collaboratively and in real time. Direct communication among technical experts reduces organizational latency and improves operational agility.

---

## 4.3 Operational Tool Development

**Key Theme:** Mission operations software and planning tools should be treated as mission-critical systems requiring continuous development, validation, and operational support throughout the mission lifecycle.

### Background

OSIRIS-REx demonstrated that modern planetary exploration depends as much on mission operations software as on spacecraft hardware. Observation planning, science sequencing, navigation, command generation, constraint verification, and mission execution relied on an integrated suite of operational tools that evolved continuously throughout the mission. These tools became essential components of the operational architecture rather than supporting infrastructure.

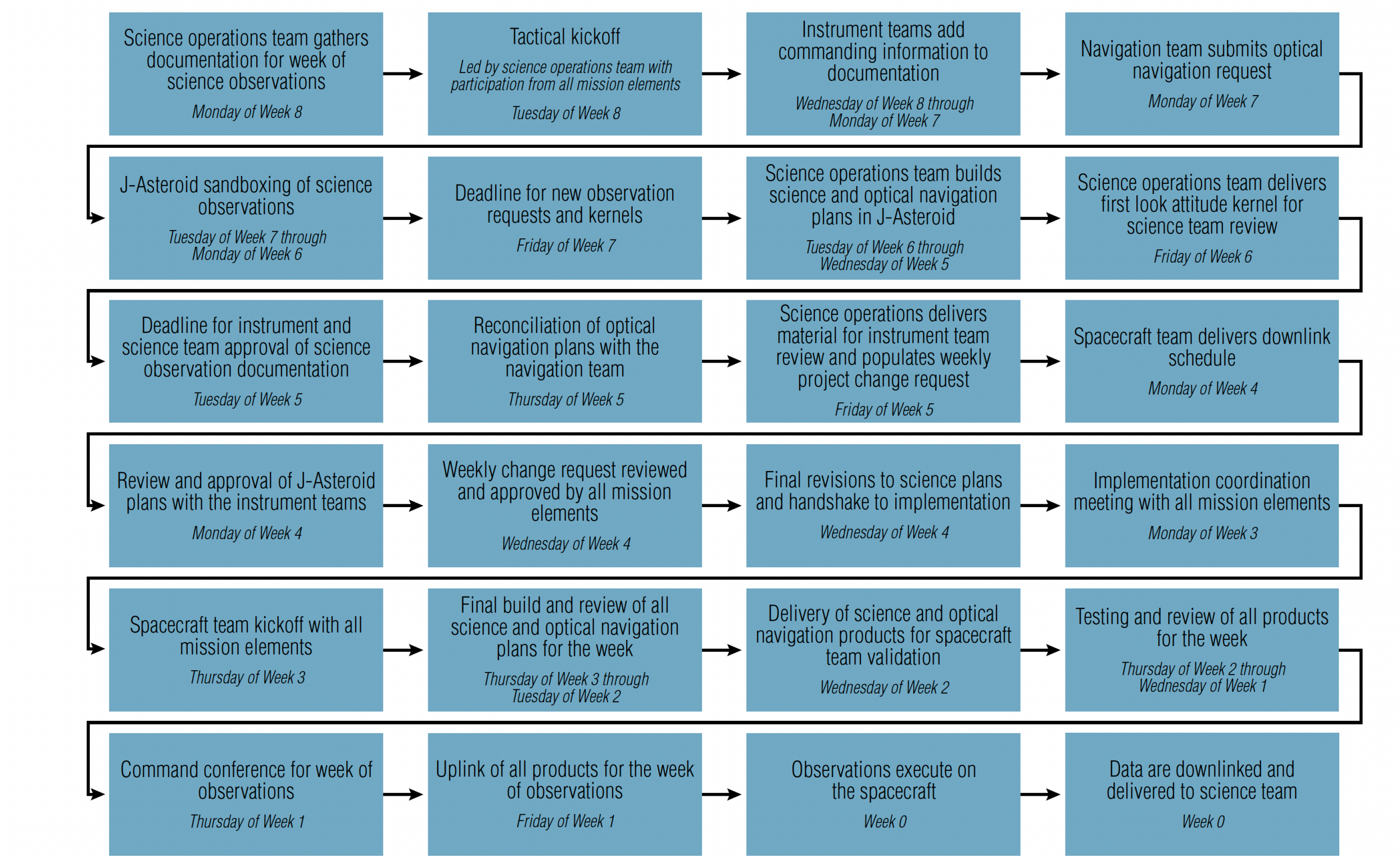

**Figure 7**. Flow diagram of the OSIRIS-REx science planning and implementation process, illustrating the eight-week timeline and the interaction between multiple mission elements. From: Lauretta et al. (2021), In *Sample return missions* (pp. 163-194). Elsevier. Credit: University of Arizona. Credit: University of Arizona.

During formulation and early development, however, the complexity of both the science planning effort and its associated software tools was significantly underestimated. Early planning assumed that operational software could largely be adapted from heritage systems, particularly the J-Mars planning environment. As the mission matured, it became clear that Bennu's unique operational environment—including small-body navigation, rapidly evolving mission geometry, dynamic observation constraints, and the inability to recover missed observations—required substantially more sophisticated capabilities than originally anticipated.

The evolution of J-Asteroid illustrates this lesson. Although it was built upon heritage software, nearly every phase of Bennu operations required new functionality, revised algorithms, additional validation, and operational refinements. Early requirements proved incomplete because many mission-specific operational challenges could only be understood after flight experience accumulated. Rather than reaching a stable end state before launch, J-Asteroid underwent continuous development throughout the mission in response to evolving operational requirements.

Science planning personnel also developed complementary tools—including J-Planner2, Activity Inspector, tactical planning scripts, standardized planning templates, and comprehensive observation sensitivity analyses—to evaluate navigation uncertainty, observation geometry, slew and settle margins, image overlap, and operational constraints. These tools enabled rapid tactical replanning while maintaining mission safety and maximizing scientific return. Because Bennu observations were frequently non-repeatable, rigorous pre-execution validation and continuous refinement of planning products became essential elements of mission success.

OSIRIS-REx revealed that operational software should be regarded as a living component of the flight system. Like spacecraft software, mission planning tools require realistic staffing, continuous systems engineering, configuration management, verification, and dedicated operational support throughout the mission lifecycle.

## Lessons Learned

- **Treat mission operations software as mission-critical flight infrastructure**. Planning tools, sequencing software, operational databases, and validation systems require the same engineering discipline, configuration management, and lifecycle support as spacecraft flight software. Core capabilities should be flight-ready, but dedicated development teams should remain in place throughout operations to incorporate new requirements, resolve anomalies, and adapt tools to evolving mission conditions.
- **Scope operational software development realistically**. Science planning, mission sequencing, navigation integration, and operational validation are often substantially more complex than anticipated during formulation, particularly for first-of-a-kind missions.
- **Design operational tools to accommodate uncertainty**. Navigation errors, evolving operational constraints, slew margins, observation geometry, and tactical replanning should be incorporated into planning architectures from the outset rather than added as later enhancements.
- **Develop flexible, modular software architectures that can evolve** throughout the mission. Adaptable tools are substantially more resilient than rigid systems when mission requirements change unexpectedly.
- **Perform rigorous pre-execution validation** through sensitivity analyses, simulations, and automated verification. For missions where observations cannot easily be repeated, early error detection and operational validation become mission-critical capabilities.

## 4.4 Operational Rehearsals and Early Flight Testing

**Key Theme:** Early flight activities should be used deliberately to validate team coherence, tools, procedures, and operational interfaces before critical mission events.

### Background

OSIRIS-REx treated early flight activities as opportunities to comprehensively rehearse mission operations under realistic flight conditions, rather than simply to execute secondary scientific objectives or demonstrate isolated engineering capabilities. The Earth Trojan Asteroid Search and Earth Gravity Assist campaigns provided valuable opportunities to exercise the operational system—including personnel, software, procedures, communications, navigation, and decision-making—well before arrival at Bennu.

These early campaigns exposed numerous deficiencies that would have been far more difficult to resolve during proximity operations. Operational rehearsals identified weaknesses in kernel management, optical navigation planning, command product generation, operational tool maturity, and cross-organizational coordination. They also revealed opportunities to improve workload distribution, refine operational procedures, strengthen communication pathways, and validate interfaces among science planning, flight dynamics, spacecraft operations, and mission management.

Because these issues were discovered during relatively low-risk mission phases, the project had sufficient time to revise procedures, improve operational software, foster team coordination, and mature mission planning processes before the significantly more demanding environment encountered at Bennu. By the time proximity operations began, the operational system—including people, tools, and processes—had already accumulated substantial flight experience.

OSIRIS-REx demonstrated that operational readiness cannot be established through ground testing alone. Real flight environments reveal interactions among spacecraft behavior, operational software, organizational processes, and human decision-making that are difficult to reproduce prior to launch. Deliberately using early mission opportunities to stress test the complete operational system substantially reduced mission risk and improved performance during critical operations.

### Lessons Learned

- **Treat early flight activities as operational rehearsals** rather than isolated engineering demonstrations or secondary science campaigns. Every early mission phase should contribute directly to operational readiness.

- **Leverage real flight conditions to exercise the complete operational system**—including personnel, planning tools, flight software, procedures, communication pathways, and organizational interfaces—before critical mission events.
- **Identify operational weaknesses early**, while sufficient schedule remains to implement corrective actions. Early discovery of deficiencies in software, procedures, staffing, or coordination significantly reduces risk during later mission phases.
- **Validate cross-functional interactions** as rigorously as individual technical systems. Operational success depends on the integrated performance of people, tools, organizations, and decision processes.
- **Capture lessons from each rehearsal** and incorporate them systematically into subsequent operations. Operational readiness should improve continuously throughout the mission rather than relying solely on pre-launch preparation.

---

## 4.5 Launch Site Safety and External Operations Coordination

**Key Theme:** Hazardous operations conducted by neighboring organizations must be incorporated into launch site risk assessments, operational planning, and personnel safety protocols.

**Background**

During final spacecraft processing, OSIRIS-REx was integrated within the Vehicle Integration Facility (VIF) at Cape Canaveral's Space Launch Complex 41 while other launch operations were simultaneously being conducted elsewhere on the Eastern Range. During this period, SpaceX was preparing a Falcon 9 vehicle at nearby Space Launch Complex 40 for a static-fire test involving a fully fueled launch vehicle. On September 1, 2016, the Falcon 9 exploded during propellant loading, producing a large fireball, debris field, and the release of hazardous combustion products.

At the time of the explosion, OSIRIS-REx personnel were working inside the Pad 41 VIF, and the Atlas V launch vehicle was simultaneously undergoing a wet dress rehearsal. Fortunately, no injuries occurred, but the event demonstrated that hazardous operations conducted by neighboring organizations can present significant risks to personnel and flight hardware well beyond the immediate launch pad.

The incident highlighted the need for launch-range safety planning to extend beyond the boundaries of individual launch complexes. Personnel safety assessments should account for adjacent hazardous operations, toxic plume dispersion, blast hazards, debris footprints, and concurrent launch processing activities across the entire launch complex. Effective coordination among launch providers, payload organizations, and range management is essential whenever hazardous operations are conducted in close proximity.

**Lessons Learned**

- **Assess launch-site hazards beyond the launch pad**, considering cumulative hazards rather than limiting risk analyses to individual launch pads or facilities. Increasing launch cadence at shared launch facilities requires corresponding improvements in cross-organizational coordination and launch-site safety planning.
- **Integrate operational risk assessment and coordinate hazardous operations** among all organizations sharing a launch range. Fueling operations, static-fire tests, wet dress rehearsals, and spacecraft processing should be evaluated collectively to identify cumulative hazards and unacceptable concurrent risks.
- **Protect personnel by establishing exclusion zones and evacuation procedures** whenever hazardous operations at adjacent facilities present credible blast, debris, or toxic exposure risks.

---

# 5. Sample Return, Curation, and Analysis

## 5.1 Sample Analysis Planning

**Key Theme:** Analysis of returned samples should be planned, budgeted, and managed as a mission phase with the same rigor applied to spacecraft development and flight operations.

### Background

Experience from OSIRIS-REx demonstrated that sample investigation should be treated as a continuation of mission operations rather than post-project science. The mission consisted of three interconnected scientific campaigns: pre-launch astronomical characterization, remote sensing investigation at Bennu, and laboratory analysis of the returned samples. Whereas the first two campaigns received dedicated project management, schedule margin, and systems engineering support, the third was originally planned as a relatively short post-mission activity.

The baseline mission allocated approximately two years for initial sample analysis. Once the Bennu samples were returned, however, it became clear that the extraordinary diversity, complexity, and scientific significance of the collection required substantially more time. New discoveries continually generated new hypotheses, additional analytical requirements, and follow-on investigations that could not have been anticipated during mission formulation. Rather than concluding the mission, sample return initiated an extended period of scientific discovery.

Sample analysis had not initially been provided with sufficient schedule margin, financial reserves, or dedicated project management to accommodate these evolving scientific opportunities. Although Key Decision Point F formally marked completion of the flight project, there was no corresponding transition into a fully managed sample-analysis phase. This gap was addressed by establishing the Sample Analysis Team under the direction of the Principal Investigator, organizing the effort into working groups with designated leads, and defining formal responsibilities for sample allocation, data-product completion, publication planning, and progress reporting. The campaign was also extended by an additional year to accommodate the longer-than-expected timeline required to open the sample collector, complete preliminary examination and cataloging, and distribute material to the analysis team. This experience has subsequently informed NASA policy by recognizing sample analysis as an integral phase of future sample-return missions.

## Lessons Learned

- **Plan sample curation and scientific investigation as a formal mission phase** rather than a post-project activity. Sample analysis is the culmination of a sample-return mission and should receive management attention comparable to spacecraft development and flight operations.
- **Provide realistic schedule margin, financial reserves, and project management support** for sample analysis. Scientific discovery inevitably generates new questions, analytical requirements, and opportunities that cannot be fully anticipated during formulation. Mission planning should preserve sufficient flexibility to pursue transformative discoveries rather than limiting investigations to those envisioned before launch.
- **Maintain continuity between flight and laboratory operations**. Knowledge developed during spacecraft operations should transition directly into sample curation and laboratory analysis through sustained project leadership, systems engineering, and science coordination.

## 5.2 Contamination Knowledge and Curation

**Key Theme:** Contamination knowledge programs must extend continuously from spacecraft development through curation, sample distribution, and scientific analysis to preserve confidence in scientific discoveries.

### Background

From its inception, OSIRIS-REx recognized that returning a pristine sample required more than minimizing contamination—it required documenting any contamination that could not be eliminated. Accordingly, contamination knowledge was elevated to a Mission Level-1 requirement alongside returning a pristine sample. The project established a comprehensive contamination control and knowledge program that began during spacecraft formulation and continued through assembly, test, launch, cruise, sample acquisition, Earth return, curation, and laboratory investigation (Dworkin et al., 2018).

The contamination knowledge program archived hundreds of hardware samples, witness plates, environmental monitors, engineering materials, and contamination reference specimens while maintaining detailed records of spacecraft materials, cleaning processes, and potential contamination sources. Contamination witness plates were exposed throughout ATLO, archived materials were preserved for future comparison, and known contamination risks were continuously documented and updated as the mission evolved.

This long-term investment proved invaluable following sample return. During preliminary sample analysis, investigators encountered numerous materials that initially appeared to be contaminants. Rather than relying on assumptions, each suspected contaminant was evaluated using archived materials, contamination records, witness samples, laboratory blanks, and structured investigative processes. In several cases, materials initially suspected to be terrestrial contamination were ultimately identified as indigenous components of Bennu. Other materials were traced to laboratory handling rather than spacecraft contamination. The contamination knowledge archive prevented both false-positive contamination claims and the inadvertent rejection of authentic extraterrestrial discoveries.

OSIRIS-REx therefore demonstrated that contamination knowledge does not end when the spacecraft launches or even when the sample returns to Earth. As analytical capabilities improve and new scientific questions emerge over decades of investigation, archived contamination information becomes an increasingly valuable scientific resource that preserves confidence in the interpretation of returned samples.

### Lessons Learned

- **Treat contamination knowledge as a mission-long scientific activity** rather than a spacecraft development task**.** Contamination knowledge protects scientific discovery as much as contamination control protects sample cleanliness.
- **Archive contamination knowledge materials—** such as spacecraft hardware, witness plates, environmental samples, and processing records—so future investigators can evaluate suspected contaminants using direct comparisons rather than assumptions.
- **Extend contamination knowledge beyond end of mission.** Laboratory blanks, contamination monitoring, and centralized documentation should accompany distributed samples throughout their scientific lifetime.
- **Maintain contamination knowledge archives for decades.** Future analytical techniques will exceed current capabilities and may require re-evaluation of contamination sources that cannot yet be anticipated.

---

## 5.3 Sample Analysis Training and Rehearsals

**Key Theme:** Scientific investigations should be rehearsed before sample return using realistic practice campaigns that train personnel, validate procedures, and exercise coordination across distributed analytical teams.

**Background**

The OSIRIS-REx Sample Analysis Science Operations Planning Exercises (SA-SOPIEs) were developed to prepare the Sample Analysis Team for the unprecedented challenge of receiving, curating, allocating, and analyzing pristine asteroid material. Rather than relying solely on written procedures, the project conducted structured practice campaigns that simulated the complete sample analysis workflow—from receipt of samples through allocation, documentation, analytical planning, data reporting, and cross-disciplinary interpretation.

These rehearsals familiarized investigators with sample handling protocols, contamination control requirements, analytical priorities, communication pathways, and decision-making processes before the arrival of the Bennu samples. Just as importantly, they established realistic expectations regarding the pace of scientific investigation, the complexity of coordinating numerous laboratories, and the need to balance rapid discovery with careful documentation and contamination assessment.

The exercises also exposed procedural weaknesses, clarified team roles and responsibilities, refined analytical workflows, and strengthened collaboration among geographically distributed investigators. By the time the samples were returned to Earth, the science team had already practiced addressing many of the logistical and organizational challenges they would encounter during the initial analysis campaign.

The experience demonstrated that scientific readiness is as important as spacecraft readiness. Sample-return missions should view scientific operations as an operational phase requiring rehearsal, refinement, and continuous improvement before precious extraterrestrial material is available.

### Lessons Learned

- **Conduct comprehensive sample analysis rehearsals** well before sample return, using realistic operational scenarios and representative materials whenever possible. Use rehearsals to identify procedural gaps, refine documentation, validate communication pathways, and clarify organizational roles before mission-critical operations begin.
- **Exercise the complete scientific workflow**—including curation, sample allocation, contamination assessment, analytical planning, data management, reporting, and cross-disciplinary interpretation—rather than focusing solely on laboratory techniques.
- **Prepare investigators for the operational realities** of coordinated sample analysis, including distributed teams, competing analytical priorities, contamination concerns, and rapid scientific decision-making.
- **Treat scientific operations readiness with the same rigor applied to spacecraft flight operations**, recognizing that the scientific return from a sample-return mission depends as much on organizational preparedness as on the successful return of the sample itself.

---

## 5.4 Mission Authority Through Sample Analysis

**Key Theme:** Sample curation, allocation, and scientific investigation should remain integrated within the mission management structure through completion of the primary science objectives.

## Background

OSIRIS-REx was conceived as a PI-led sample-return mission in which spacecraft development, flight operations, and sample acquisition, curation, and analysis formed a single integrated mission. From the earliest project planning, curation was established as a mission element reporting within the overall mission management structure alongside spacecraft, science, and operations. This organization ensured that decisions affecting sample handling, contamination knowledge, preliminary examination, sample allocation, and scientific priorities remained closely coordinated with the mission's Level-1 scientific objectives (Righter et al., 2023).

This integrated approach worked effectively throughout mission development and sample return. However, the exceptionally long duration of sample-return missions creates organizational challenges that extend beyond the flight project itself. During OSIRIS-REx, substantial turnover occurred within supporting NASA organizations responsible for curation and program oversight. As institutional knowledge of the original mission management philosophy diminished, differing interpretations emerged regarding roles, responsibilities, and authority for post-return activities. These differing perspectives complicated coordination during the transition from sample return to long-term scientific investigation.

The experience demonstrated that returning the sample is not the completion of the mission. Sample curation, allocation, and analysis are direct extensions of mission execution and remain essential to achieving the mission's scientific objectives. Maintaining an integrated management structure through completion of the primary science campaign preserves accountability, ensures consistent scientific priorities, and minimizes organizational friction during this critical phase.

## Lessons Learned

- **Treat sample curation as an integral mission function** rather than a separate institutional activity. Curation decisions directly affect the scientific return of the mission and should remain aligned with mission objectives.
- **Extend PI authority and responsibility** through preliminary examination, sample allocation, and completion of the primary science investigation. Responsibility for mission success should remain matched with the authority needed to achieve it.
- **Anticipate substantial personnel turnover** for long-duration sample-return missions, within both the project and supporting NASA organizations. Clear governance documentation, formal transition planning, and preservation of institutional memory are essential to maintaining continuity.

- **Document roles, responsibilities, and decision authority early** for curation, sample allocation, database management, contamination knowledge, and scientific oversight. Reaffirm them throughout the mission lifecycle, particularly following significant organizational or leadership transitions.
- **Maintain a unified management structure** from mission conception through completion of the primary science campaign. Continuity reduces organizational conflict and maximizes the scientific return from irreplaceable extraterrestrial samples.

---

# 6. Data Archiving

## 6.1 Spacecraft Data Archiving and Long-Term Scientific Stewardship

**Key Theme:** Scientific data archives should be treated as enduring mission products. Early planning, close collaboration with archive partners, standardized production pipelines, and flexibility to accommodate evolving data standards are essential for maximizing long-term scientific return.

### Background

The scientific legacy of a planetary mission depends not only on returning samples and collecting observations, but also on preserving those data in forms that remain discoverable, understandable, and reusable for decades. From the beginning of OSIRIS-REx, the Science Data Management Plan established the Planetary Data System (PDS) as the permanent archive for all mission science products and defined a schedule for delivering calibrated observations, derived products, and ancillary datasets.

OSIRIS-REx was one of NASA's earliest missions to fully adopt the PDS4 archive standard. This required close collaboration with the Small Bodies Node of the PDS throughout mission development to ensure that archive products satisfied evolving technical standards while remaining scientifically useful. Early engagement with PDS personnel proved invaluable as both the mission and the archive system matured together.

For standard instrument products, the project developed a configuration-controlled processing pipeline that generated all raw-to-calibrated data products. Because the pipeline itself underwent comprehensive peer review, subsequent archive deliveries required only verification of incremental products rather than repeated reviews of calibration methodology. This approach substantially reduced review time for each delivery while maintaining archive quality. Success depended on maintaining close communication with PDS whenever pipeline modifications were proposed.

The mission also partnered directly with the Navigation and Ancillary Information Facility (NAIF) to produce and archive all SPICE kernels. Leveraging the expertise of the organization responsible for SPICE standards improved archive quality while reducing project cost and schedule burden.

Derived science products proved substantially more challenging than anticipated. Although specifications for many products were established before launch, the final products evolved considerably as Bennu observations revealed new scientific opportunities and operational realities. Each unique derived product type required development of new PDS4 labels and metadata, and the effort required to generate these archive products significantly exceeded early estimates. The complexity arose not from data volume, but from the diversity of products and the extensive metadata required to preserve their scientific utility.

The mission also experienced significant changes in NASA's broader data management environment over its 15-year lifecycle. During that period, NASA transitioned from the PDS3 to the PDS4 standard, implemented Science Policy Documents SPD-41 and SPD-41A, adopted Digital Object Identifiers (DOIs) as standard practice, and expanded data availability requirements for scientific publications. While these developments ultimately strengthened scientific reproducibility and data accessibility, they also required substantial adaptation by a long-duration mission that had begun under a different set of archival expectations.

**Lessons Learned**

- **Treat scientific data archives as primary mission deliverables** rather than end-of-mission documentation activities**.** Archive planning, staffing, and resources should be integrated into mission development from the outset.
- **Establish close, continuous collaboration with the Planetary Data System and archive partners** early in mission development. Frequent communication allows archive products and standards to evolve together and helps avoid costly late-stage redesign.
- **Develop configuration-controlled processing pipelines** for standard data products whenever possible. Validated pipelines simplify subsequent archive deliveries, reduce review effort, and improve consistency across the mission.

- **Leverage specialized organizations for highly technical archive products.** Partnering directly with experts such as NAIF for SPICE products improves quality while reducing project cost and schedule risk.
- **Allocate substantially more time and resources for derived data products than for standard calibrated products.** Metadata development, archive labeling, documentation, and evolving scientific products often become the dominant archival effort.
- **Expect data management policies and archive standards to evolve** during long-duration missions. Archive plans should be sufficiently flexible to accommodate new standards, metadata requirements, persistent identifiers, and publication policies without disrupting mission schedules.
- **View data stewardship as a permanent component of the mission's scientific legacy.** Well-documented, discoverable, and reusable archives maximize the long-term value of mission investments and enable scientific investigations far beyond the lifetime of the flight project.

---

## 6.2 Sample Analysis Data Management and Sample Information Stewardship

Key Theme: Data systems supporting large, multi-technique sample-analysis programs must combine standardized requirements with sufficient flexibility to accommodate evolving analytical plans, new techniques, and changing scientific understanding. Open data formats, adaptable metadata standards, and integrated sample tracking are essential for preserving the long-term scientific value of returned samples.

### Background

The scientific value of returned samples depends not only on their physical preservation, but also on maintaining reliable records of sample provenance, analytical history, laboratory measurements, and relationships among subdivided materials. For the OSIRIS-REx Sample Analysis Team, the Sample Analysis Micro-Information System (SAMIS) was developed to support these functions across a large, geographically distributed, multi-institutional research program.

One of the principal lessons from SAMIS was that data standards must be both consistent and adaptable. The system needed to establish common requirements across a wide range of analytical techniques while accommodating substantial differences in instrument capabilities, native data structures, laboratory practices, and scientific objectives. Standards that were too rigid would have been unable to support newly identified analytical needs, whereas insufficient standardization would have produced inconsistent and difficult-to-integrate records.

Proprietary instrument formats presented a significant challenge for long-term data access. Requiring investigators to deliver data in open formats improved the likelihood that analytical results would remain accessible independent of particular instruments, vendors, or software packages. However, conversion from proprietary formats could impose a substantial burden on investigators and did not always preserve the full set of instrument metadata. In some cases, supplementary PDF/A documents or text files were therefore required to retain information that could not be incorporated directly into the open-format data product.

Full interoperability with other geochemical and astromaterials databases also remained difficult. The astromaterials community does not yet have broadly endorsed, machine-readable ontologies or persistent identifier systems for many analytical techniques and laboratory instruments. The absence of these community standards limited the ability to create fully interoperable records within SAMIS and demonstrated that some data-management challenges require coordination beyond the scope of any individual mission or project.

The Sample Analysis Plan and the composition and activities of the Sample Analysis Team evolved continuously as the returned material was characterized and new scientific questions emerged. SAMIS therefore required mechanisms for updating standards while preserving backwards compatibility. The ability to add technique definitions, incorporate optional metadata, and accept supplementary files allowed the system to respond to emerging needs without invalidating previously submitted records.

SAMIS was particularly successful as a sample-tracking and sample-creation system. It provided a consistent mechanism for recording sample locations and transfers, preserving parent–child relationships among subdivided samples, enforcing required naming conventions, and collecting metadata needed for eventual delivery to the sample catalog. Embedding these requirements directly within the system reduced the likelihood of duplicate, incomplete, or incorrectly named records and improved consistency across the distributed analysis team.

## Lessons Learned

- **Treat sample-information management as a core component of the returned-sample science program**. Sample provenance, analytical history, location, transfer records, and subdivision relationships should be captured systematically from the beginning of sample analysis.

- **Establish common data and metadata requirements across analytical techniques**, Design those standards to accommodate differences among instruments, laboratories, and scientific objectives.
- **Require open, nonproprietary data formats whenever practical to support long-term accessibility**. At the same time, recognize that conversion may be burdensome and that supplementary files may be necessary to preserve instrument-specific metadata.
- **Plan explicitly for information that cannot be represented within a standardized analytical data file**. PDF/A documents, text files, and other supplementary records may be needed to retain calibration details, instrument configurations, processing histories, or other contextual information.
- **Build flexibility and backwards compatibility into sample-analysis information systems**. Standards, technique definitions, and metadata requirements will evolve as the sample is characterized and new analytical priorities emerge.
- **Allow optional metadata and supplementary materials where appropriate**. A tiered approach can preserve required consistency while enabling investigators to document technique-specific or newly identified information.
- **Integrate sample creation, naming, transfer, and location tracking directly into the data-management system**. Automated enforcement of naming conventions and parent–child relationships reduces duplicate, incorrect, and incomplete records.
- **Coordinate the development of analytical ontologies and persistent identifiers at the broader community level**. Individual missions can implement interim solutions, but full interoperability among astromaterials and geochemical databases will require sustained community-wide standards development.
- **Recognize that sample stewardship and data stewardship are inseparable**. The long-term scientific value of returned material depends on preserving both the physical samples and the information needed to understand their provenance, processing history, and analytical context.

---

# Summary & Conclusions

## 1. Leadership and Governance

**Unified leadership enables disciplined execution.** A PI-led management structure that integrates science, engineering, project planning, and systems engineering under a common decision-making framework provides the discipline needed to maintain alignment among mission objectives, technical execution, cost, schedule, and risk throughout the mission lifecycle. [Section 1.1]

**A Management Advisory Council during Phase A enabled efficient cross-institutional support**. It is important for the MAC to have clearly defined membership, charter, and authority. Convene it regularly throughout the mission—not only during crises—to maintain executive engagement and institutional alignment. [Section 1.2]

**Mission continuity depends on more than stable leadership.** Long-duration missions succeed when continuity is preserved across personnel, institutional knowledge, management processes, and mission planning. A living Design Reference Mission, disciplined succession planning, and protected schedule margin provide the resilience needed to adapt to new discoveries while maintaining alignment with mission objectives. [Section 1.3]

**Authority, contracts, and partnerships must work together.** Successful PI-led missions align organizational authority, contractual relationships, and acquisition strategies with mission responsibilities. Effective governance provides the flexibility to manage risk, resolve issues rapidly, and preserve mission-critical capabilities as technical and programmatic conditions evolve. [Section 1.4]

## 2. Cost, Schedule, and Risk Management

**Risk management is about ensuring mission success—not preserving every capability.** Effective projects distinguish between capabilities that can be removed with acceptable impact and those whose long-term scientific or operational value justifies additional investment. The discipline to make both decisions, using objective criteria established before crises emerge, is a hallmark of successful mission management. [Section 2.1]

**Budgets are built by engineers—not accountants.** The most credible mission cost estimates emerge when scientists, engineers, project managers, and planners develop budgets together from a common technical baseline. Integrated cost planning improves transparency, strengthens accountability, and reduces the likelihood of costly surprises later in the project. [Section 2.2]

**Reserves should buy down risk—not simply cover overruns.** Projects that establish financial reserves and schedule margin early gain the flexibility to solve problems before they become crises, preserving mission objectives while maintaining technical and programmatic discipline. [Section 2.3]

**Discovery requires time.** Exploration missions should allow scientific knowledge—not predetermined schedules—to drive critical decisions. Protecting schedule margin during both flight operations and sample analysis enables informed decision-making, reduces mission risk, and maximizes long-term scientific return. [Section 2.4]

## 3. Systems Engineering and Technical Management

**Interfaces are engineered systems.** On planetary missions, the greatest risks often arise not within individual subsystems but at the interfaces between organizations, disciplines, software, hardware, and operational processes. Successful projects devote the same rigor to defining interface ownership, authority, communication pathways, and integration planning as they do to designing flight hardware. Investing in interface management early reduces downstream coordination costs, accelerates decision-making, and improves mission resilience. [Section 3.1]

**Design for discovery, not certainty.** Every planetary exploration mission encounters environments that differ from pre-launch expectations. Systems engineered for flexibility—through modular architectures, operational margin, adaptable software, and expandable interfaces—can respond to new knowledge without compromising mission success. Adaptability is therefore not a contingency capability but a core design principle for exploration missions. [Section 3.2]

**Ground systems are flight systems.** Modern planetary missions depend as much on planning software, data-processing pipelines, mission operations centers, and science planning infrastructure as they do on spacecraft hardware. Underestimating ground system complexity leads to rework, operational inefficiency, and increased mission risk. Projects should apply equivalent systems engineering discipline, staffing, and management attention to both flight and ground segments throughout the mission lifecycle. [Section 3.3]

**Resilience comes from capabilities, not just hardware.** Traditional redundancy duplicates components to protect against failure. Functional redundancy instead exploits overlapping capabilities among different systems, allowing existing hardware to perform new roles when mission conditions change. A thorough understanding of system capabilities can transform apparent single-point failures into manageable operational challenges. [Section 3.4]

**Heritage is knowledge, not drawings.** The greatest value of heritage missions lies not simply in reusing components, but in preserving the engineering understanding that explains why those components perform successfully. When that knowledge is lost, projects risk inheriting documentation without inheriting the experience needed to interpret it. Independent verification and layered redundancy remain essential, even for proven flight systems. [Section 3.5]

## 4. Mission Operations

**Operations succeed at the speed of communication.** During high-tempo operations, organizational delays can become mission risks. Co-locating technical experts and fostering trusted cross-functional relationships enable rapid assessment, informed decision-making, and coordinated responses to evolving operational conditions. For complex exploration missions, organizational integration is a mission capability in its own right. [Section 4.1]

**Trust is built before the crisis.** Operational teams cannot create transparency, mutual confidence, and shared understanding during a high-pressure event. These capabilities must be developed deliberately through face-to-face engagement, visible tools, clear decision authority, and sustained cross-functional collaboration before critical operations begin. [Section 4.2]

**Mission operations software is never finished.** Exploration missions inevitably encounter operational conditions that cannot be fully anticipated during development. Core capabilities must be validated before launch, but planning software, sequencing tools, operational databases, and analysis systems require continuous refinement as mission knowledge evolves. Projects should budget for, staff, and manage operational software as a living engineering system rather than assuming development ends at launch. [Section 4.3]

**Practice in flight before performance matters.** Ground simulations cannot fully reproduce the operational complexity of a spacecraft operating in space. Early mission activities provide invaluable opportunities to validate the complete operational system under real flight conditions, allowing teams to identify weaknesses, improve tools, refine procedures, and build confidence before the mission enters its highest-risk phases. [Section 4.4]

**Launch-site safety extends beyond the launch pad.** Modern spaceports support multiple providers operating simultaneously within a shared infrastructure. Safe mission execution requires coordinated risk management across neighboring facilities so that hazardous operations conducted by one organization do not unnecessarily endanger personnel, spacecraft, or infrastructure supporting another mission. [Section 4.5]

## 5. Sample Return, Curation, and Analysis

**Returning the sample is the beginning of the final mission campaign.** The ultimate scientific value of a sample-return mission is realized through decades of laboratory investigation. Project planning, budgeting, and management should recognize sample analysis as an operational phase of the mission, supported by dedicated resources, schedule flexibility, and sustained scientific leadership, rather than treating it as post-project research. [Section 5.1]

**Cleanliness alone is not enough.** Even the cleanest sample-return mission will encounter unexpected materials during scientific investigation. The scientific value of a returned sample depends not only on preventing contamination but also on preserving the knowledge needed to distinguish terrestrial contamination from authentic extraterrestrial discoveries. OSIRIS-REx showed that contamination knowledge is a permanent scientific asset that should remain available throughout the lifetime of the returned collection. [Section 5.2]

**Rehearse the science—not just the spacecraft.** The SA-SOPIE exercises demonstrated that practicing scientific operations before sample return improves technical readiness, strengthens collaboration, and allows procedural issues to be resolved before irreplaceable samples become available. Future sample-return missions should invest in scientific rehearsal campaigns with the same discipline applied to mission operations rehearsals (SOPIEs). [Section 5.3]

**Sample curation is part of the mission.** The scientific investigation of the returned material is the culmination of the mission, not a separate activity. The organizational structure that successfully delivers the sample should remain intact through curation, sample allocation, and completion of the primary science objectives to preserve accountability and maximize scientific return. [Section 5.4]

## 6. Data Archiving

**The spacecraft data archive is the mission's permanent legacy.** Spacecraft eventually cease operations and samples are gradually consumed through scientific investigation, but properly curated scientific data remain available indefinitely. Investing early in robust archive development, standardized data products, comprehensive metadata, and long-term stewardship ensures that future generations of researchers can continue to extract new discoveries from mission datasets long after flight operations have ended. [Section 6.1]

**Build sample science data systems around consistent standards** while preserving the flexibility to add new techniques, metadata, and supplementary files as the science evolves. Open formats, backwards compatibility, and integrated sample tracking are essential to the long-term value of returned samples [Section 6.2].